\documentclass[aps,twocolumn,preprintnumbers,amsmath,amssymb]{revtex4}

\usepackage{graphicx}
\usepackage{dcolumn}
\usepackage{bm}

\begin{document}

\preprint{CERN-PH-TH/2005-097}

\title{Regular two-component bouncing cosmologies and perturbations therein}

\author{V. Bozza$^{a,b,c}$ and G. Veneziano$^{d,e}$}
\affiliation{ $^a$ Centro Studi e Ricerche ``Enrico Fermi'', via
Panisperna 89/A, Rome, Italy. \\
 $^b$ Dipartimento di Fisica ``E. R. Caianiello", Universit\`a di
Salerno, I-84081 Baronissi, Italy. \\
 $^c$ Istituto Nazionale di Fisica Nucleare, Sezione di Napoli,
 Naples, Italy.\\
 $^d$ CERN, PH Dept., TH Unit, CH-1211 Geneva 23, Switzerland.\\
 $^e$ Coll\`ege de France, 11 place Marcelin Berthelot, F-75005 Paris, France.
}

\date{\today}

\begin{abstract}
We present a full investigation of scalar perturbations in a
rather generic model for a regular bouncing universe, where the
bounce is triggered by an effective perfect fluid with negative
energy density. Long before and after the bounce the universe is
dominated by a source with positive energy density, which may be a
perfect fluid, a scalar field, or any other source with an
intrinsic isocurvature perturbation. Within this framework, we
present an analytical method to accurately estimate the spectrum
of large-scale scalar perturbations until their reentry, long
after the bounce. We also propose a simple way to identify
non-singular gauge-invariant variables through the bounce and
present the results of extensive numerical tests in several
possible realizations of the scenario. In no case do we find that
the spectrum of the pre-bounce growing mode of the Bardeen
potential can be transferred to a post-bounce constant mode.
\end{abstract}


\maketitle

\section{Introduction}

Motivations for studying bouncing cosmologies come from today's
widespread belief that a consistent theory of quantum gravity
should cure the initial cosmological singularity --as well as
other singularities that plague General Relativity (GR)--  by a
suitable modification of its short-distance behaviour. In the
context of string theory, bouncing cosmologies of different kinds
emerge as the most reasonable alternative to the Big Bang
singularity. In the pre-bounce phase, as one goes to
past-infinity, the universe is driven towards a non-singular
attractor where generic initial conditions can be chosen.

The first proposal of a bouncing string cosmology was the
so-called Pre-Big Bang scenario \cite{PBB}, which exploits the
non-minimal coupling of the dilaton in superstring theory to drive
a Pre-Big Bang super-inflationary epoch, ending when higher-order
derivatives and/or loop corrections become non-negligible. A
second realization of the same idea was given by the
ekpyrotic/cyclic scenario \cite{Ekp}, inspired by Horawa--Witten
braneworlds \cite{HorWit}. Here the pre-bounce is characterized by
the slow approach of two parallel branes, which eventually collide
and then move away, giving rise to an ordinary expanding universe
on each of the two branes. Both scenarios, when viewed in a
four-dimensional Einstein frame, can be represented by a universe
bouncing from contraction to a standard expansion.

A crucial test for any cosmological model lies on its capability
of accounting for the origin, the growth and the spectrum of the
fluctuations needed  to explain the large-scale structures and the
anisotropies in the cosmic microwave background. It is well known
that observations require a nearly scale-invariant spectrum for
the Bardeen potential. Such a spectrum  can be naturally provided
by the initial quantum fluctuations of the inflaton in the
standard inflationary picture \cite{MFB}. Alternatively, it can be
provided by the initial isocurvature perturbations of another
field, as in the ``curvaton''  mechanism \cite{Curv}, which is
naturally implemented through  the axion field in the Pre-Big Bang
scenario \cite{Axion}. A third possibility, the claim that the
ekpyrotic scenario could directlty provide (without appealing to
the curvaton mechanism) a scale-invariant spectrum of adiabatic
density perturbations, has given rise to a considerable debate
about the correct way to describe the effects of a bounce on
cosmological perturbations \cite{Others}.

The main point at issue here concerns the Bardeen potential: it
certainly grows very large, with a red or scale-invariant
spectrum, during the pre-bounce phase. Does this mode survive the
bounce or does it decay away, allowing other modes to take over?
The use of classical matching conditions for perturbation through
a space-like hypersurface \cite{DerMuk} (which assume the
continuity of the comoving curvature $\zeta$) is not trivial in
the case of a bounce, since the Hubble rate changes sign and the
null energy condition (NEC) is necessarily violated for some time
during the bounce. Then, more exotic matching conditions cannot be
excluded, depending on the physics of the bounce \cite{DurVer}.

In this situation, the only possibility to test some general arguments
is to construct explicit models of regular bounces where the final
spectrum of the Bardeen potential can be explicitly evaluated. In
the context of GR, the more conservative possibility is to study
closed universes, where the bounce is possible without violating
the NEC \cite{K=1}. In this case, the presence of the curvature
scale is responsible for an explicit  momentum dependence
 in the transfer matrix from pre- to post-bounce perturbations.

On the other hand, if one wants to stick to spatially flat
cosmologies, a very good approximation presumably after a long
pre-bounce epoch, one has to introduce a negative energy-density
component that determines the transition from a contracting
universe to an expanding one, dominated by a standard component,
such as radiation. In this respect, Peter \& Pinto-Neto
\cite{PetPin} studied a bounce induced by a ghost scalar field,
which connects a radiation-dominated contraction to a
radiation-dominated expansion. Surprisingly they found that the
Bardeen potential keeps the spectrum of its pre-bounce growing
mode. Similar conclusions have been found by Finelli \cite{Fin} in
a model with two perfect fluids (one with negative energy
density).

On the other hand, explicit bounce examples have been built using
only one scalar field and triggering the bounce by higher
derivatives and loop corrections to the Einstein equations
\cite{Car} or by potentials involving integrals on spatially flat
sections of a compact universe \cite{GGV}. Both these
one-component examples pointed at a complete post-bounce decay of
the Bardeen pre-bounce growing mode.

Coming back to two-component models, Allen \& Wands \cite{AllWan}
found again a decay of the Bardeen pre-bounce growing mode in a
cosmology dominated by a scalar field with a tracking potential
(which allows it to mimic the background evolution of any kind of
fluids), with the bounce triggered by a ghost scalar field. The
reasons why the conclusions of this work are in contrast with
those of Ref. \cite{PetPin} remained unclear.

In this paper we would like to bring more clarity in the
investigation of regular bouncing cosmologies by introducing a
rather generic cosmological model, which includes, as particular
cases in a much wider parameter space, all previous models
containing two non-interacting components.  We find that the
pre-bounce growing mode of the Bardeen potential always matches a
decaying mode in the post-bounce, leaving no trace in the final
spectrum at horizon re-entry. Our analysis proves that the claims
in Refs. \cite{PetPin,Fin} are incorrect, while it confirms the
results of Ref. \cite{AllWan}, generalizing to a much larger class
of two-component bouncing cosmologies. A short summary of our
results was anticipated in Ref. \cite{BozVen}.

In this work we provide all the details of the analysis
summarized in Ref. \cite{BozVen}. In particular, we explicitly
provide  the steps necessary to produce the analytical estimates
from the method sketched in Ref. \cite{BozVen}. Moreover, we
present the numerical studies of the cosmological perturbations in
the whole parameter space, testing and supporting the analytical
estimates. Finally, we expand the discussion on the gauge-choice
issue and the way to construct non-singular gauge-invariant
variables.

In Sect. 2, we introduce the class of backgrounds we want to
consider, write down all the equations, and discuss the general
behaviour of our background quantities. In Sect. 3, we introduce
scalar perturbations, discuss the gauge issue, introduce the most
convenient gauge-invariant quantities for our analysis,  and
discuss the vacuum normalization of perturbations that  provides
the initial conditions for our analysis. In Sect. 4, we describe
the method used for  analytical estimates and  provide all the
necessary steps for a systematic approach of the problem. In Sect.
5, we discuss these estimates throughout the parameter space that
describes our bouncing cosmologies. In Sect. 6, we numerically
check all the estimates discussed in Sect. 5. Finally, Sect. 7
contains a discussion of the results.

\section{A general model for two-component bouncing cosmologies}

A common criticism against bouncing cosmologies triggered by a
ghost scalar field --or negative energy matter--  is that they
deal with unphysical sources that are plagued by quantum
instabilities and/or inconsistencies. However, we must keep in
mind that the ``true"  bounce should be the outcome of a
short-distance modification of GR, such as those coming from
String/M-theory. Once we choose to perform an effective
description in the Einstein frame, we cannot avoid (in a spatially
flat cosmology) violations of  the NEC in the sense that the
short-distance corrections must enter as an effective negative
energy (or pressure) contributions to the right-hand side of
Friedman's equations. Unless this is  accepted, one must give up
using the tools of GR and wait for the full String/M-theory to
tell us how scalar perturbations  behave during the high-energy
phase.

In this paper we assume that a General Relativity description is
indeed possible (at least for modes that are far outside the
horizon during the bounce) and that the high-energy corrections
can be modelled as an effective new negative-energy density field
$\rho_b$, which becomes negligible as soon as we go far from the
bounce itself. At this point it makes no sense to worry about the
quantum instabilities of such an effective field, since the
quantum description of high-energy corrections can only be done in
the full mother theory. Nevertheless, nothing prevents the
investigation of classical perturbations with the standard tools
of GR.

The class of models we investigate in this paper have the following
characteristics:

\begin{itemize}
\item The universe is filled by two components $a$ and $b$, with
energy density $\rho_a>0$ and $\rho_b<0$, respectively.

\item Far from the bounce, $\rho_a$ dominates the evolution of the
universe, while $\rho_b$ is important only at the bounce and
becomes subdominant elsewhere.

\item The two components do not decay and interact only
gravitationally.

\item Far from the bounce, the ratios
\begin{equation}
\Gamma_m=p'_m/\rho'_m, ~~~~~m=a,b,
\end{equation}
are asymptotically constant, that is we have a stable attractor in
the past (implying also $w_m=p_m/\rho_m=\Gamma_m$).

\end{itemize}

The Friedman equations for a two-component cosmology can be simply
written as
\begin{eqnarray}
&& 3\mathcal{H}^2=a^2 \rho \\ && \mathcal{H}^2+2\mathcal{H}'=-a^2
p
\\ && \rho'+3\mathcal{H}(\rho+p)=0, \label{Eqrho'}
\end{eqnarray}
where a prime denotes the derivative w.r.t. conformal time $\eta$,
$a$ is the scale factor, $\mathcal{H}=a'/a$ is the conformal
Hubble rate, and we are using units such that $8\pi G=1$. The
total energy density is the sum of the individual energy densities
of the two cosmological components $\rho=\rho_a+\rho_b$, and the
same happens for the pressure $p=p_a+p_b$. Since the two fluids
have neither interactions nor decay, they satisfy the continuity
equation (\ref{Eqrho'}) independently.

As a consequence of the constancy of $\Gamma_m$, the attractor in
the past is described by the following power laws
\begin{eqnarray}
&& a(\eta)\simeq |\eta|^q \label{PowLawa}\\
&& \mathcal{H}\simeq\frac{q}{\eta} \\
&& \rho_a\simeq \frac{3q^2}{|\eta|^{2+2q}} \\
&& \rho_b\simeq -\frac{3q^2}{|\eta|^{3q(1+\Gamma_b)}} \\
&& q=\frac{2}{1+3\Gamma_a}. \label{PowLawq}
\end{eqnarray}
Here we have already chosen a convenient normalization for the
scale factor and for the conformal time $\eta$, so that the two
fluids have comparable energy densities at $\eta\simeq -1$. Of
course, the condition $\rho_b/\rho_a\rightarrow 0$ for $\eta
\rightarrow -\infty$ is satisfied if and only if
$\Gamma_b>\Gamma_a$.

\begin{figure}
\resizebox{\hsize}{!}{\includegraphics{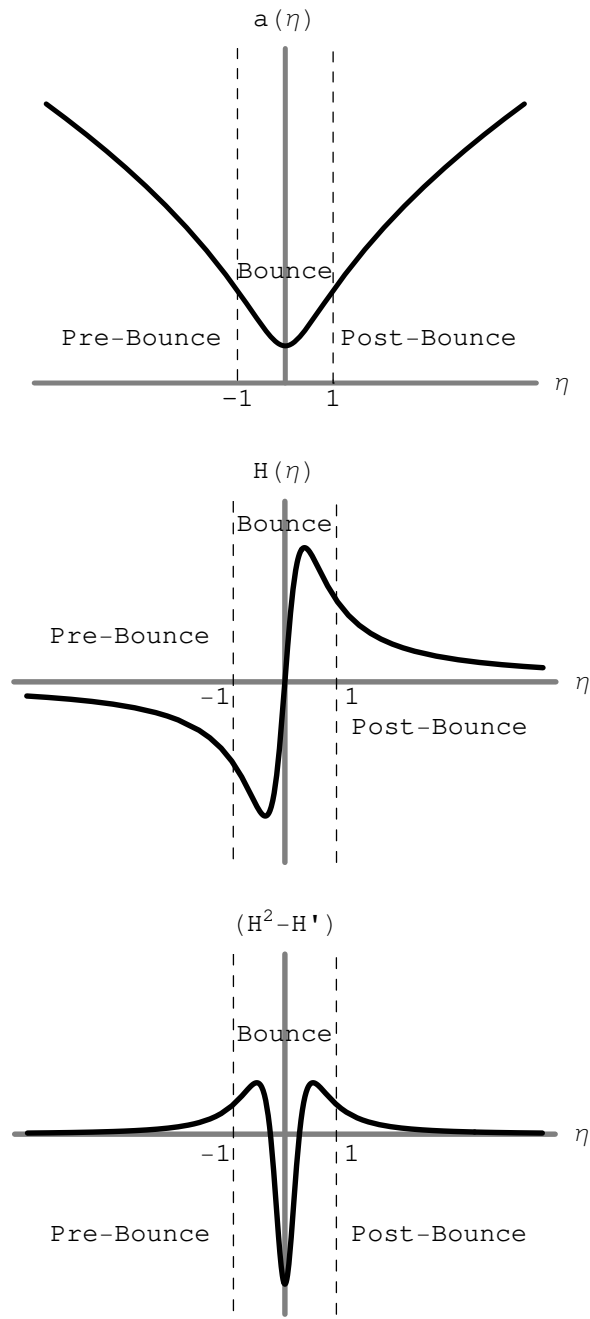}}
\caption{Typical behaviours of the background functions in a
bouncing cosmology. At the top we have the scale factor, in the
middle the conformal Hubble parameter $\mathcal{H}$, and at the
bottom the function $(\mathcal{H}^2-\mathcal{H}') =a^2(\rho+p)/2$,
which becomes negative when NEC violation occurs. The borders of
the bounce phase have been put in evidence by vertical dashed
lines at the times $\eta=\pm 1$.
 }
 \label{FigBkgd}
\end{figure}

The power-law solution holds until $\eta\sim-1$. Then the bounce
occurs at $\eta=0$, when $\mathcal{H}$ changes sign and finally at
some time $\eta\sim 1$ the post-bounce expanding power-law
solution is established with the same $\Gamma_a$ and $\Gamma_b$ as
in the pre-bounce, so that Eqs. (\ref{PowLawa})-(\ref{PowLawq})
hold again. In practice, we allow the bounce to be asymmetric
within the range $|\eta| \lesssim 1$, but require the asymptotic
power laws to be the same before and after the bounce. Actually,
this final assumption could be easily relaxed, but we do not know
any explicit example of two-components regular bounce with
different pre- and post-bounce power laws. In Fig. \ref{FigBkgd}
we show typical behaviours of the background functions.

\section{Scalar perturbations}

The perturbed line element, including only scalar perturbations,
is
\begin{eqnarray}
&ds^2=&a^2(\eta)\left\{ (1+2\phi) d\eta^2-2B_{,i} d\eta dx^i
\right.
\nonumber \\
&& \left. -\left[(1-2\psi)\delta_{ij}+2E_{,ij}\right]dx^i dx^j
\right\},
\end{eqnarray}
while the perturbed energy-momentum tensor of each source reads:
\begin{equation}
{{(T_m)}^\mu}_\nu= \left(%
\begin{array}{cc}
  \rho_m+\delta \rho_m & -(\rho_m+p_a) \mathcal{V}_{m,j} \\
  (\rho_m+p_m) \mathcal{V}_{m,i} & -(p_m+\delta p_m) \delta_{ij}+ \Xi_{m,ij} \\
\end{array}%
\right) \label{TmunuP}
\end{equation}
with $m=a,b$ respectively; $\mathcal{V}_m$ is the velocity
potential and $\Xi_m$ is the anisotropic stress. It is worth
recalling the effects of a gauge transformation on each of these
variables. For an infinitesimal coordinate change
\begin{equation}
\eta\rightarrow \tilde{\eta}=\eta+\xi^0, ~~~~ x^i \rightarrow
\tilde{x}^i=x^i+\delta^{ij}\xi_{,j},
\end{equation}
the scalar perturbations are modified in the following way:
\begin{eqnarray}
&& \tilde{\phi}=\phi-\mathcal{H}\xi^0-{\xi^0}' \label{Gaugephi}\\
&& \tilde{\psi}=\psi+\mathcal{H}\xi^0 \\
&& \tilde{B}=B+\xi^0-\xi' \\
&& \tilde{E}=E-\xi \\
&& \widetilde{\delta \rho}_i=\delta \rho_i-\rho_i'\xi^0 \\
&& \widetilde{\delta p}_i=\delta p_i-p_i'\xi^0 \\
&& \tilde{\mathcal{V}}_i=\mathcal{V}_i-\xi^0. \label{GaugeVi}
\end{eqnarray}

\subsection{A general model for source perturbations}

In order to close the system of equations, for each component we
need to specify a relation between the perturbative variables.
Since we want to make a single treatment for both perfect fluids
and scalar fields, we need a  general enough relation, containing
both kinds of sources as particular cases. We therefore impose,
for the dominant source,
\begin{equation}
\delta p_a = c_a^2 \delta \rho_a + \alpha (\rho_a + p_a)
\mathcal{V}_a, ~~~~~~~  \alpha = \frac{p_a'-c_a^2
\rho_a'}{\rho_a+p_a}, \label{EOS}
\end{equation}
while for the secondary component we choose perfect fluid
relations
\begin{equation}
\delta p_b = c_b^2 \delta \rho_b, \;\;  c_b^2=\Gamma_b.
\label{EOSb}
\end{equation}
We also assume the anisotropic stress $\Xi_m$ to vanish for both
components.

In general, we allow $\Gamma_a$ and $c_a^2$ to be different,
opening the possibility of a non-vanishing $\alpha$. In order to
better understand the physical meaning of $c_a^2$ and $\alpha$ we
can go to the  comoving gauge with respect to the dominant fluid
($\mathcal{V}_a=0$). We then have
\begin{equation}
c_a^2=\frac{\delta p_a|_{\mathcal{V}_a=0}}{\delta
\rho_a|_{\mathcal{V}_a=0}},
\end{equation}
which identifies $c_a$ as the sound speed of the $a$-fluid. In
fact, this quantity is properly defined in the frame where the
fluid is at rest. We will see later that, indeed, $c_a$ is the
velocity of propagation of fluctuations in the $a$-fluid, thus
explaining our somewhat unconventional notation.

The perfect fluid relation is recovered when the equality
$\Gamma_a=c_a^2$ holds. In fact, in this case, we simply have
$\alpha=0$ and the pressure perturbation is just proportional to
the density perturbation in any gauge with the same coefficient
$\Gamma_a=c_a^2$. On the other hand, if $\Gamma_a \neq c_a^2$, we
have $\alpha \neq 0$, implying the presence of an intrinsic
isocurvature mode
\begin{equation}
\tau_a \delta S_a = \delta p_a- (p_a'/\rho_a')\delta
\rho_a=\alpha(\rho_a + p_a)\left(\mathcal{V}_a-\delta
\rho_a/\rho_a'\right). \label{Iso}
\end{equation}

In particular, a scalar field $\varphi$ with a self-interaction
potential $V(\varphi)$ has
\begin{eqnarray}
&& \rho_a= \frac{{\varphi'}^2}{2a^2}+V(\varphi), \;\;\; p_a=
\frac{{\varphi'}^2}{2a^2}-V(\varphi),  \\ && \delta \rho_a=
\frac{\varphi'}{a^2}\delta \varphi'-\phi\frac{{\varphi'}^2}{a^2}
+V'(\varphi) \delta \varphi, \;\;\;  \mathcal{V}_a= \frac{\delta
\varphi}{\varphi'}, \\ && \delta p_a= \frac{\varphi'}{a^2}\delta
\varphi'-\phi\frac{{\varphi'}^2}{a^2}  -V'(\varphi) \delta
\varphi,
\end{eqnarray}
while the covariant conservation of the energy-momentum tensor
gives
\begin{equation}
\varphi''+2\mathcal{H}\varphi'+a^2 V'(\varphi)=0.
\end{equation}
It is immediate to prove that Eq. (\ref{EOS}) holds with $c_a^2=1$
and $\alpha=-2a^2V_{,\varphi}/\varphi'$. Notice that a free scalar
field is completely equivalent to a perfect fluid with stiff
equation of state, $p = \rho$, while the presence of any potential inevitably
introduces an intrinsic isocurvature mode.

From this discussion, we conclude that Eq. (\ref{EOS}) is general
enough to contain perfect fluids, scalar fields and any source
with an intrinsic isocurvature mode generated according to Eq.
(\ref{Iso}). This ensures our treatment of a wide generality and,
in particular, it allows us to include all previously studied
models as particular cases.

\subsection{The gauge issue}

Sooner or later, all authors studying the problem of the evolution
of cosmological perturbations through the bounce have met a
non-trivial difficulty in making a convenient choice of the gauge.
The commonly used longitudinal gauge makes the Bardeen potential
appear directly as one entry of the perturbed  metric. Typically,
this variable becomes very large in the approach to the bounce,
making the validity of the perturbative approach questionable.
Other commonly used gauges introduce explicit divergences at the
NEC violation or at the bounce. Under some circumstances, a safe
gauge is provided by setting to zero the perturbation of a scalar
field whose time derivative never changes sign \cite{GGV,AllWan}.

Our approach is to assume the existence of a regular gauge where
all the components of the metric and of the energy-momentum
tensors stay finite. In order for the linear  theory to be valid,
the perturbations should also stay small. If no regular gauge
exists, then we have to conclude that the background is not a
stable solution of Einstein equations and thus it is uninteresting
as a cosmological model. Our approach will be applicable to all
regular bounces that are stable solutions of the Einstein
equations. In practical terms, a regular gauge must have $\phi$,
$\psi$, $B_{,i}$, $E_{,ij}$, $\delta \rho_m$, $\delta p_m$ and
$(\rho_m+p_m)\mathcal{V}_m$ finite and small ($m=a,b$). Notice
that it is the combination $(\rho_m+p_m)\mathcal{V}_m$ that enters
the energy-momentum tensor of the $m$-source. This means that
$\mathcal{V}_m$ can possibly diverge when $(\rho_m+p_m)$ goes to
zero provided the above product stays finite. The same happens for
the total velocity potential:
\begin{equation}
\mathcal{V}=(\rho+p)^{-1}\sum\limits_{m=a,b}(\rho_m+p_m)\mathcal{V}_m,
\end{equation}
which may diverge at the NEC violation, the product
$(\rho+p)\mathcal{V}$ staying finite.

Using the regular gauge, it is easy to construct gauge-invariant
combinations that are regular and can thus be safely used to
follow perturbations through the bounce. Indeed, the Bardeen
potential
\begin{equation}
\Psi=\psi+\mathcal{H}(E'-B) \label{Bardeen}
\end{equation}
is a regular combination of regular perturbations. Old matching
conditions suggest to use the comoving curvature perturbation
$\zeta=\psi+\mathcal{H}\mathcal{V}$. However, the presence of the
velocity potential does not protect this variable from possible
divergences. We propose to use instead
\begin{equation}
\tilde{\zeta}=(\mathcal{H}^2-\mathcal{H}')(\psi+\mathcal{H}\mathcal{V}),
\end{equation}
which is guaranteed to be regular, since
$2(\mathcal{H}^2-\mathcal{H}')\mathcal{V}=a^2(\rho+p)\mathcal{V}$
is finite in the regular gauge. We then introduce individual
gauge-invariant variables
\begin{eqnarray}
&&\tilde{\zeta}_m= \frac{1}{2}a^2(\rho_m+p_m) (\psi+\mathcal{H}\mathcal{V}_m ) \label{Defzetai}\\
&& \Psi_m= \frac{1}{2} a^2 (\rho_m' \psi+ \mathcal{H}\delta \rho_m
),\label{DefPsii}
\end{eqnarray}
which are also regular by construction. Of course
\begin{equation}
\tilde{\zeta}=\tilde{\zeta_a}+\tilde{\zeta_b}.
\end{equation}

Before proceeding with  the perturbation equations, it is worth
commenting briefly  on the commonly used gauges. Using the
transformations (\ref{Gaugephi})--(\ref{GaugeVi}), we can pass
from the (assumed) regular gauge to any other gauge. These
transformations may either explicitly induce divergences, or
ensure that the variables in the transformed gauge are regular as
well. Let us analyse some cases explicitly.

The zero-curvature gauge ($\tilde{\psi}=0$), obtained by the
regular gauge setting $\xi=-\psi\mathcal{H}^{-1}$, introduces
terms with $\mathcal{H}^{-1}$ in $\phi$, $B$ and $\mathcal{V}_i$,
which diverge at the bounce.

The comoving gauge ($\tilde{\mathcal{V}}=0$), obtained by setting
$\xi^0=\mathcal{V}$, introduces $\mathcal{V}$ in all other
variables. As said before, $\mathcal{V}$ can diverge at the NEC
violation.

The synchronous gauge ($\tilde{\phi}=\tilde{B}=0$) is obtained by
setting $\xi^0=a^{-1}\int a \phi d\eta$ and $\xi=\int Bd\eta+\int
d\eta a^{-1}\int a\phi d\eta'$. This transformation induces no
divergences and so we can say that if a regular gauge exists where
all perturbations are finite, then  the variables in the
synchronous gauge are also finite.

Similarly, for the longitudinal gauge ($\tilde{E}=\tilde{B}=0$),
obtained by setting $\xi=E$ and $\xi^0= B-E'$, no divergence is
introduced.

The uniform density gauge ($\tilde{\delta \rho}=0$) is obtained by
setting $\xi^0=\delta \rho/\rho'$. By the continuity equation, we
see that the transformed variables have to diverge both at the NEC
violation and at the bounce $\mathcal{H}=0$.

Concerning a possible ``uniform pressure gauge" ($\tilde{\delta p}=0$, obtained
by setting $\xi^0=\delta p/p'$) it is difficult to say something
general. However, it is hard to imagine a model where $p'$ does
not vanish at any time during the bounce.

The uniform individual density or pressure gauges have a similar
fate. They typically introduce divergences at the bounce. The
gauge comoving with one of the components, instead, introduces
divergences only if $\mathcal{V}_m$ itself diverges, but this is
not necessarily true. The regularity of the gauges in Refs
\cite{GGV,Car,AllWan} can be understood in the light of this
remark.

Among those just described, the synchronous and longitudinal
gauges are the only ones that keep all the variables finite. Yet
we know that in the longitudinal gauge $\psi=\Psi$ and the Bardeen
potential grows very large in the approach to the bounce. Indeed,
the finiteness of all variables does not assure their smallness.
In this case, one can still consider the variables in the
longitudinal gauge as useful auxiliary mathematical quantities:
the transformation to longitudinal-gauge variables has to be
regarded as a simple mathematical change of variables, without the
physical meaning of a coordinate transformation. Nevertheless,
nothing prevents one from solving the equations in the
longitudinal-gauge variables and then discuss the results with
gauge-invariant quantities, whose physical interpretation is
clear. In fact, the longitudinal variables have the important
property of being regular (even if large) and thus can be used to
go through the bounce safely.

\subsection{Perturbation equations}

Let us start by perturbing the Einstein equations. The 00, $i0$,
$ii$, $i\neq j$ components respectively read
\begin{eqnarray}
&& \!\!\!\!\!\!\!\! a^2 \delta \rho +6\mathcal{H}(\mathcal{H} \phi
+\psi')+2k^2[\psi+\mathcal{H}(E'-B)]=0 \\
&& \!\!\!\!\!\!\!\!2(\mathcal{H}\phi +\psi')=a^2(\rho+p)\mathcal{V} \label{i0}\\
&& \!\!\!\!\!\!\!\!-\frac{1}{2}a^2 \delta p+ \mathcal{H}^2 \phi
+2\mathcal{H}'\phi
+\mathcal{H} \phi' +2\mathcal{H} \psi' + \psi'' =0 \\
&&\!\!\!\!\!\!\!\! \psi-\phi+2\mathcal{H}(E'-B)+(E''-B')=0.
\end{eqnarray}

We can use Eq. (\ref{i0}) to eliminate $\phi$, Eqs. (\ref{EOS})
and (\ref{EOSb}) for $\delta p_m$, Eqs. (\ref{Defzetai}) and
(\ref{DefPsii}) to eliminate $\delta \rho_m$ and $\mathcal{V}_m$,
and finally Eq. (\ref{Bardeen}) for $\psi$. Equations 00, $ii$ and
$i\neq j$ become
\begin{eqnarray}
&& \!\!\!\!\!\!\!\! (3\tilde{\zeta}+k^2 \Psi)\mathcal{H}+ \Psi_a+\Psi_b = 0 \label{00}\\
&& \!\!\!\!\!\!\!\! \tilde{\zeta}'=
(\alpha -\mathcal{H})\tilde{\zeta}_a-\mathcal{H}\tilde{\zeta}_b+ c_a^2\Psi_a+c_b^2\Psi_b \label{ii}\\
&&\!\!\!\!\!\!\!\!
\Psi'+\frac{2\mathcal{H}-\mathcal{H}'}{\mathcal{H}^2}\Psi=\frac{\tilde{\zeta}}{\mathcal{H}},
\label{ij}
\end{eqnarray}
where we have made use of the fact that
$\tilde{\zeta}_a+\tilde{\zeta}_b=\tilde{\zeta}$.

The perturbations of the covariant conservation of the
energy-momentum tensor of the first fluid, after using the
Hamiltonian constraint (\ref{00}), give
\begin{eqnarray}
& \tilde{\zeta}_a'&
+\frac{2\mathcal{H}(\mathcal{H}-\alpha)+a^2(\rho_b+p_b)}{2\mathcal{H}}\tilde{\zeta}_a
-c_a^2 \Psi_a = \nonumber \\
&& =\frac{a^2(\rho_a+p_a)}{2\mathcal{H}}\tilde{\zeta}_b  \label{Eqza}\\
& \Psi_a'& +\frac{6c_a^2
\mathcal{H}^2+a^2(\rho+p+\rho_a+p_a)}{2\mathcal{H}}\Psi_a
+a^2\frac{\rho_a+p_a}{2\mathcal{H}}\Psi_b
\nonumber \\
&+&\!\!\!\!
\frac{2k^2+6\mathcal{H}\alpha+3a^2(\rho_a+p_a)}{2}\tilde{\zeta}_a=-3a^2\frac{\rho_a+p_a}{2}
\tilde{\zeta}_b \label{EqPsia}.
\end{eqnarray}
The perturbations of the covariant conservation of the
energy-momentum tensor of the second fluid give the same equations
with $a \leftrightarrows b$. The sum of Eq. (\ref{Eqza}) with its
$b$ counterpart yields Eq. (\ref{ii}), while the sum of Eq.
(\ref{EqPsia}) with its counterpart yields a linear combination of Eqs. (\ref{ij}) and
 (\ref{ii}).

Summing up, Eqs. (\ref{Eqza}) and (\ref{EqPsia}) and their $b$
counterparts are an independent set of coupled first-order
differential equations, which completely describe cosmological
perturbations in our problem in terms of the regular
gauge-invariant variables $\tilde{\zeta}_a$, $\tilde{\zeta}_b$,
$\Psi_a$ and $\Psi_b$. The fact that $\mathcal{H}$ appears in some
denominators does not indicate that the differential equations
break down at the bounce. Actually, one can easily check by the
definitions (\ref{Defzetai}) and (\ref{DefPsii}) that the
numerators also contain an $\mathcal{H}$ factor multiplied by
finite functions.

Later on, we shall widely use another set of variables to describe
the system, namely $\tilde{\zeta}$, $\tilde{\zeta}_b$, $\Psi$ and
$\Psi_b$. This set has the advantage of including directly $\Psi$
and $\tilde{\zeta}$, which are the most interesting quantities
characterizing cosmological perturbations. In terms of these
variables the set of differential equations becomes
\begin{eqnarray}
& \!\!\!\! \!\!\!\!\Psi'& \!\!\!\!
+\frac{2\mathcal{H}^2-\mathcal{H}'}
{\mathcal{H}}\Psi-\frac{\tilde{\zeta}}{\mathcal{H}}=0
\label{EqPsi} \\ \nonumber \\
& \!\!\!\! \!\!\!\!\tilde{\zeta}'&
\!\!\!\! + (\mathcal{H}+3c_a^2\mathcal{H} -\alpha)\tilde{\zeta}
+c_a^2 k^2 \mathcal{H}\Psi =-\alpha\tilde{\zeta}_b \nonumber \\ &&
+(c_b^2-c_a^2)\Psi_b \\ \nonumber \\
& \!\!\!\! \!\!\!\!\Psi_b'& \!\!\!\! +
\frac{\mathcal{H}^2(1+3c_b^2)-\mathcal{H}'}{\mathcal{H}} \Psi_b
+k^2 \tilde{\zeta}_b = \nonumber \\
&& = \frac{a^2 (\rho_b +p_b)}{2}  k^2\Psi \\ \nonumber \\
& \!\!\!\! \!\!\!\!\tilde{\zeta}_b'& \!\!\!\!
+\frac{2\mathcal{H}^2-\mathcal{H}' }{\mathcal{H}} \tilde{\zeta}_b
-c_b^2 \Psi_b =\frac{a^2(\rho_b+p_b)}{2\mathcal{H}} \tilde{\zeta}.
\label{Eqzb}
\end{eqnarray}

\subsection{Vacuum normalization} \label{Sect. vacuum}

Choosing initial conditions for the perturbations of the two
components is by no means trivial. While for the dominant
component we can choose ordinary vacuum normalization, for the
secondary one this is quite an arbitrary choice. In fact, we can
think that the first component is an ordinary perfect fluid, a
scalar field, or any other field initially in a vacuum state. On
the contrary, on the basis of the former discussion about the
nature of the secondary component, we cannot really justify giving
a well-defined quantum initial state to this effective component.
Yet, in order
 to compare our results with other works, which have
systematically chosen the vacuum normalization even for the
secondary component, we will follow this attitude as well. In the
conclusions we will comment about the generality of our results
and their (in)dependence on the initial state of the
secondary-component perturbations.

Taking the derivatives of Eq. (\ref{Eqza}) and eliminating
$\Psi_a$ and $\Psi_a'$ using Eqs. (\ref{Eqza}) and (\ref{EqPsia}),
we get a second-order equation for $\tilde{\zeta}_a$ coupled to
$\tilde{\zeta}_b$ and $\Psi_b$. Of course, we also get an
analogous second-order equation for $\tilde{\zeta}_b$. Introducing
the Sasaki--Mukhanov canonical variables
\begin{equation}
v_{m}= \tilde{\zeta}_m/[c_m (\rho_m+p_m)^{1/2}\mathcal{H}],
\label{DefvMi}
\end{equation}
the two equations can be written in a very compact way as
\begin{eqnarray}
&& v_{a}''+\left(c_a^2k^2-z_a''/z_a \right)v_{a} = O(\rho_b/\rho_a)^{1/2} \label{vMa}\\
&& v_{b}''+\left(c_b^2k^2-z_b''/z_b \right)v_{b} =
O(\rho_b/\rho_a)^{1/2}, \label{vMb}
\end{eqnarray}
with $z_a=a^2(\rho_a+ p_a)^{1/2}/(c_a \mathcal{H})$ and
$z_b=a^{(1+3c_b^2)}(\rho_b+ p_b)^{1/2}/c_b$. Inside
$O(\rho_b/\rho_a)^{1/2}$ we include all the terms where $v_a$,
$v_b$ appear multiplied by background functions going to zero
asymptotically faster than $|\eta|^{-2}$ and terms where $v_a'$,
$v_b'$ are multiplied by functions going to zero faster than
$|\eta|^{-1}$. All these terms become irrelevant in the asymptotic
past and do not affect the initial vacuum normalization. We also
see that $c_m^2$ multiplies the Laplacian of $v_m$ and thus
represents the velocity of propagation of the fluctuations in each
component, confirming our statement in Sect. 2.

The asymptotic solutions of Eqs. (\ref{vMa}) and (\ref{vMb}) in
the power law regime are
\begin{eqnarray}
&  v_{a}=& \sqrt{|\eta|}C_a H^{(1)}_{\nu_a}(c_a k
|\eta|) \label{Hankela}\\
&  v_{b}=& \sqrt{|\eta|} C_b H^{(1)}_{\nu_b}(c_b k
|\eta|),\label{Hankelb}
\end{eqnarray}
where $H_\nu^{(1)}$ is the Hankel function of the first kind,
$C_a$ and $C_b$ are pure numbers of order 1, and
\begin{eqnarray}
&&\nu_a=\frac{1}{2}-q \label{nua}\\
&& \nu_b=\frac{1}{2}(3c_b^2q-1-q). \label{nub}
\end{eqnarray}

Now that we have established the initial conditions for the
Mukhanov variables, we can derive those for $\tilde{\zeta}_a$ and
$\tilde{\zeta}_b$ from Eq. (\ref{DefvMi}). For $\Psi_a$ and
$\Psi_b$ we can use Eq. (\ref{Eqza}) and the corresponding
equation for $\tilde{\zeta}_b'$.

\section{Evolution of perturbations}

In this section we shall estimate the behaviour of perturbations
before and after the bounce, giving a complete discussion of the
final momentum and time dependences in terms of the parameters
of the cosmological model.

The first step is to rewrite Eqs. (\ref{EqPsi})--(\ref{Eqzb}) in
their integral form
\begin{eqnarray}
& \!\!\!\!\!\!\!\! \Psi & \!\!\!\! =
 \frac{\mathcal{H}}{a^2}
\left[ \frac{c_1(k)}{k^2}+ \int \frac{a^2}{\mathcal{H}^2}
\tilde{\zeta} d\eta \right], \label{Psiint}
\\ \nonumber \\
& \!\!\!\!\!\!\!\! \tilde{\zeta} & \!\!\!\! = a^2(\rho_a+p_a)
\left[ c_2(k) -\int \frac{c_a^2\mathcal{H}}{a^2(\rho_a+p_a)} k^2
\Psi d\eta \right.
\nonumber \\
&& \left. -\int \frac{\alpha}{a^2(\rho_a+p_a)} \tilde{\zeta}_b
d\eta + \int \frac{c_b^2-c_a^2}{a^2(\rho_a+p_a)}\Psi_b d\eta
\right], \label{zetaint}
\\ \nonumber \\
& \!\!\!\!\!\!\!\! \Psi_b & \!\!\!\! =
\frac{\mathcal{H}}{a^{1+3c_b^2}} \left[ c_3(k) -\int
\frac{a^{1+3c_b^2}k^2}{\mathcal{H}} \tilde{\zeta}_b d\eta \right.
\nonumber \\
&& \left.  + \int \frac{a^{3+3c_b^2}(\rho_b +
p_b)}{2\mathcal{H}}k^2 \Psi d\eta \right], \label{Psibint}\\
\nonumber \\
& \!\!\!\!\!\!\!\!\tilde{\zeta}_b & \!\!\!\! =
a^{1+3c_b^2}\mathcal{H} (\rho_b +p_b) \left[ c_4(k)+\int
\frac{c_b^2}{a^{1+3c_b^2}\mathcal{H} (\rho_b +p_b)} \Psi_b d\eta
\right.
\nonumber \\
&& \left. + \int \frac{a^{1-3c_b^2}}{2\mathcal{H}^2}
 \tilde{\zeta} d\eta \right]. \label{zbint}
\end{eqnarray}

In these equations, each variable is expressed as the sum of the
solution of the associated homogeneous differential equation,
carrying an arbitrary constant $c_i(k)$, plus several integrals of
the other variables, representing the coupling terms. This
structure is particularly useful to study perturbations outside
the horizon as the ones we are interested in. Indeed we shall
limit out attention to modes with $k \ll 1$, i.e. to modes which
exit the horizon long before the bounce and reenter much later.
For these modes, horizon-exit can still be described by the
asymptotic decoupled solutions (\ref{Hankela}) and
(\ref{Hankelb}).

As a result, we can match the expansions
(\ref{Psiint})--(\ref{zbint}) to the long-wavelength limit of the
asymptotic solutions. As a practical example, let us examine
$\tilde{\zeta}$, which far from the bounce is dominated by
$\tilde{\zeta}_a$. Through Eq. (\ref{Hankela}) and (\ref{DefvMi}),
the vacuum normalization provides the following expression in the
long-wavelength limit for this variable
\begin{equation}
\tilde{\zeta}_a \sim k^{-\nu_a}|\eta|^{-2} + k^{\nu_a}
|\eta|^{-1-2q}, \label{zmatchvacuum}
\end{equation}
where all the background functions have been replaced by their
power-law behaviours (\ref{PowLawa})--(\ref{PowLawq}).

Equation (\ref{zetaint}) in the power-law limit gives
\begin{equation}
\tilde{\zeta} \sim |\eta|^{-2}\left[ c_2(k) + c_1(k)
|\eta|^{1-2q}+c_0(k) \right], \label{zmatchint}
\end{equation}
where we have used Eq. (\ref{Psiint}) to evaluate the integral of
$\Psi$ and we have kept only the lowest-order terms in $k\eta$.
The integration constant $c_0(k)$ can be simply absorbed in the
definition of $c_2(k)$.

At this point, in order to match the two expressions (\ref{zmatchvacuum})
and (\ref{zmatchint}), it is sufficient to impose
\begin{equation}
c_1(k) \sim k^{\nu_a}, \; c_2(k) \sim  k^{-\nu_a},
\end{equation}
where we have discarded any numerical factor of order 1, as we
shall do throughout our analysis. In the same way, matching
$\tilde{\zeta}_b$, we obtain
\begin{equation}
c_3(k) \sim k^{-\nu_b}, \; c_4(k) \sim  k^{\nu_b}.
\end{equation}

Defining
$\vec{F}(\eta)=\left(\Psi,\tilde{\zeta},\Psi_b,\tilde{\zeta}_b\right)$,
we can write the system (\ref{Psiint})--(\ref{zbint}) in the
compact form
\begin{equation}
\vec{F}(\eta)=\vec{F}^{(0)}(\eta)+B(\eta)\int A(\eta)
\vec{F}(\eta) d\eta, \label{vecint}
\end{equation}
where
\begin{eqnarray}
&\vec{F}^{(0)}(\eta)&= \left(
 \frac{\mathcal{H}}{a^2}
\frac{c_1(k)}{k^2}, a^2(\rho_a+p_a) c_2(k), \right. \nonumber \\
&&\left. \frac{\mathcal{H}}{a^{1+3c_b^2}} c_3(k),
a^{1+3c_b^2}\mathcal{H} (\rho_b +p_b) c_4(k) \right)
\end{eqnarray}
is the vector of the homogeneous solutions, $B(\eta)$ and
$A(\eta)$ are two matrices that can be easily read from Eqs.
(\ref{Psiint})--(\ref{zbint}). All these quantities are just
combinations of the background functions. Then we can construct
the full solution of Eq. (\ref{vecint}) recursively
\begin{eqnarray}
&&\vec{F}(\eta)=\vec{F}^{(0)}(\eta)+\sum\limits_{i=1}^\infty
\vec{F}^{(i)}(\eta) \label{vecsol} \\
&& \vec{F}^{(i)}(\eta)= B(\eta)\int A(\eta) \vec{F}^{(i-1)}(\eta)
d\eta.
\end{eqnarray}
The whole problem is thus reduced to the calculation of a
sufficient number of terms in the sum (\ref{vecsol}), which are
reduced to multiple integrals of background functions. In the
following subsections we shall illustrate how to calculate these
integrals in the pre-bounce, during the Bounce and after the
Bounce.

\subsection{Pre-bounce evolution}

As long as the power-law behaviours
(\ref{PowLawa})--(\ref{PowLawq}) remain valid, the evaluation of
all integrals can be easily done. In fact, a generic integrand
$f(\eta)$ can be expanded as $c_i(k)|\eta|^s$ giving
\begin{equation}
\int\limits^\eta c_i(k)|\eta|^s d\eta \sim c_i(k) |\eta|^{s+1},
\;\;\; \eta<-1. \label{IntPreBounce}
\end{equation}

Calculating iteratively the integrals, we can safely stop at
$\vec{F}^{(2)}(\eta)$ since the next terms would just give higher
and higher powers of $k|\eta|$ multiplying already existing terms
(recall that we are interested in the evolution of modes outside
the horizon). Keeping now the lowest order in $k\eta$ of each
mode, we find
\begin{eqnarray}
& \Psi & \sim k^{\nu_a-2} |\eta|^{-1-2q} +k^{-\nu_a} +\nonumber \\
&& k^{-\nu_b} |\eta|^{2-q(1+3c_b^2)}+C_a
k^{\nu_b}|\eta|^{1-2q}  \label{PsiPre}\\
& \tilde{\zeta}& \sim k^{\nu_a} |\eta|^{-1-2q}+k^{-\nu_a}|\eta|^{-2}+ \nonumber \\
&& k^{-\nu_b}|\eta|^{-q(1+3c_b^2)}+C_a k^{\nu_b}
|\eta|^{-1-2q} \label{zPre}\\
& \Psi_b & \sim k^{\nu_a} |\eta|^{-3q(1+c_b^2)}+k^{-\nu_a+2}|\eta|^{1-q(1+3c_b^2)}\nonumber \\
&& k^{-\nu_b} |\eta|^{-1-q(1+3c_b^2)}+k^{\nu_b+2}
|\eta|^{2-2q} \\
& \tilde{\zeta}_b & \sim k^{\nu_a}|\eta|^{1-3q(1+c_b^2)}+k^{-\nu_a}|\eta|^{-q(1+3c_b^2)}\nonumber \\
&& k^{-\nu_b} |\eta|^{-q(1+3c_b^2)}+k^{\nu_b} |\eta|^{-1-2q}.
\label{zbPre}
\end{eqnarray}

In this expression and in the following subsections, we introduce
the factors
\begin{eqnarray}
&&C_{a}=\left\{
\begin{array}{ll}
k^2\eta^2 & \;\; \alpha=0 \\
1 & \;\; \alpha \neq 0
\end{array}
\right. \\
&&X_{a}=\left\{
\begin{array}{ll}
k^2 & \;\; \alpha=0 \\
1 & \;\; \alpha \neq 0
\end{array}
\right.
\end{eqnarray}
in order to write a single expression for the solutions that holds
both when $\alpha \neq 0$ and when $\alpha=0$.

In Eq. (\ref{PsiPre}), we explicitly see the presence of a growing
mode in $\Psi$ carrying a red spectrum that becomes
scale-invariant in the limit of slow contraction $q\rightarrow 0$.
These pre-bounce solutions are valid until the energy density of
the secondary fluid becomes comparable with the dominant one. At
that point the power-law solution breaks down and the bounce phase
starts. As we have fixed this time to $\eta=-1$, by evaluating
Eqs. (\ref{PsiPre})--(\ref{zbPre}) at $\eta\simeq -1$, we obtain
the order of magnitude of the different modes at the onset of the
bounce. Since the only scale left in this evaluation is $k$, we
deduce that the mode with the reddest spectrum always dominates at
the bounce. This does not mean that the same mode will also
dominate in the post-bounce, since, as we shall see later, many of
these modes decay after the bounce.

\subsection{The Bounce}

We are not very interested in the specific evolution of all
variables during the bounce. It is just interesting to observe
that all homogeneous solutions $\vec{F}^{(0)}(\eta)$ can be easily
followed through the bounce as they are directly expressed by
combinations of background functions.

One thing that may worry the reader is the fact that some
integrals contain diverging functions of the background. Actually
the $\mathcal{H}^{-1}$ in the integrals of $\Psi$ and
$\tilde{\zeta}_b$ in Eq. (\ref{Psibint}) and in the integral of
$\Psi_b$ in Eq. (\ref{zbint}) are immediately compensated by
$\mathcal{H}$ factors in $\Psi$, $\tilde{\zeta}_b$ and $\Psi_b$
inside the integrals. The solution is a bit more subtle for the
integrals of $\tilde{\zeta}$ in Eqs. (\ref{Psiint}) and
(\ref{zbint}). For example, the integral of $\tilde{\zeta}$ in Eq.
(\ref{Psiint}) contains a $\mathcal{H}^{-2}$, which diverges as
$|\eta|^{-2}$. The integral thus diverges as $|\eta|^{-1}$, but
this divergence is compensated by the $\mathcal{H}$ factor outside
the brackets. The result is that the limit of $\Psi$ as $\eta
\rightarrow 0^-$ is finite. The same happens for the integral of
$\tilde{\zeta}$ in (\ref{zbint}). Then Eqs.
(\ref{Psiint})--(\ref{zbint}) confirm the regularity of all
variables through the bounce.

As we shall see later, we are interested in the integrals covering
the whole bounce phase
\begin{equation}
\int\limits_{-1}^{1}f(\eta)d\eta. \label{intbounce}
\end{equation}
The $k$-dependence factors out of the integral, and we are also
integrating out the $\eta$ dependence. We deduce that, in the
absence of other scales that can enter the result, this integral
must return a value with the same order of magnitude of $f(-1)$,
which has been evaluated for all functions in the previous
subsection using the pre-bounce power-law limit.

For the two integrals that diverge at the bounce, we can replace
the integral (\ref{intbounce}) by its principal value
\begin{equation}
\lim\limits_{\epsilon\rightarrow 0}\left[
\int\limits_{-1}^{-\epsilon}f(\eta)d\eta+\int\limits^{1}_{\epsilon}f(\eta)d\eta\right].
\end{equation}
This expression is consistent with the continuity of the
gauge-invariant variables. As the two divergences compensate, the
dimensional argument can be safely applied as before.

\subsection{Post-bounce evolution}

In the post-bounce, the evaluation of the homogeneous solutions
$\vec{F}^{(0)}(\eta)$ is elementary, since they are directly
expressed in terms of the background functions. In particular, we
may notice that the pre-bounce growing mode of $\Psi$ with the red
spectrum $c_1(k)k^{-2}$ comes from the homogeneous solution, which
becomes a decaying mode in the post-bounce. This does not
automatically mean that this mode becomes irrelevant in the
post-bounce. We need to show that no growing mode develops the
same spectrum and that there is always at least one mode that
dominates on this decaying mode at horizon re-entry.

In order  to evaluate the integrals in the post-bounce, we split
their integration domain into three pieces, respectively covering
the pre-bounce, the Bounce and the post-bounce phase
\begin{equation}
\int\limits^\eta
f(\eta)d\eta=\int\limits^{-1}f(\eta)d\eta+\int\limits^{1}_{-1}f(\eta)d\eta+\int\limits_{1}^\eta
 f(\eta)d\eta.
\end{equation}
These integrals give
\begin{eqnarray}
&& \int\limits^{-1}_{} c_i(k)|\eta|^s d\eta  \sim c_i(k), ~~~~
\int\limits_{-1}^{1} f(\eta) d\eta \sim c_i(k), \nonumber
\\&&
 \int\limits_{1}^{\eta} c_i(k)|\eta|^s d\eta \sim
c_i(k)\left[\eta^{s+1} -1 \right].
\end{eqnarray}
In fact, the first integral is Eq. (\ref{IntPreBounce}) evaluated
at $\eta=-1$. The second integral is Eq. (\ref{intbounce}), which
gives a result of the same order of magnitude of the pre-bounce
contribution. The last integral involves background functions in
the post-bounce and can be safely evaluated using the power-law
limit (\ref{PowLawa})--(\ref{PowLawq}) again. Summing up, each
integral gives a constant contribution plus a growing or decaying
one, depending on the value of $s$. If $s<-1$ the integral is
dominated by the constant contribution, which comes from regions
close to the bounce and from the bounce itself, while if $s>-1$
the integral is dominated by the growing contribution at late
times, far from the bounce. Note that any constant contribution,
once generated, sums to the original homogeneous mode and follows
the same evolution.

\subsection{Post-bounce solution in a simplified example} \label{Sect. simple}

In the former subsections we have explained how to calculate each
single integral once the integrand has been specified. To
construct the full solution, we need to proceed step by step,
evaluating the integrals recursively. To explain this procedure,
it is better to start with a simplified system.

Let us consider the subsystem $\Psi$, $\tilde{\zeta}$, without
couplings to the secondary component. The starting point for the
construction of the full post-bounce solution is given by the
homogeneous solutions. We thus set
\begin{eqnarray}
&&\Psi^{(0)}=\frac{\mathcal{H}}{a^2}\frac{c_1(k)}{k^2} \sim
c_1(k)k^{-2}\eta^{-1-2q} \\
&&\tilde{\zeta}^{(0)}=a^2(\rho_a+p_a) c_2(k) \sim c_2(k)\eta^{-2}.
\end{eqnarray}
As said before, the homogeneous modes survive unchanged during the
bounce and after the bounce.

The first step is to evaluate the integrals using the homogeneous
modes
\begin{eqnarray}
&\Psi^{(1)}=&\frac{\mathcal{H}}{a^2} \int
\frac{a^2}{\mathcal{H}^2} \tilde{\zeta}^{(0)} d\eta \sim \nonumber
\\
&&
c_2(k)\eta^{-1-2q} \left[ \eta^{1+2q}- 1  \right] \\
&\tilde{\zeta}^{(1)}=&- a^2(\rho_a+p_a) \int
\frac{c_a^2\mathcal{H}}{a^2(\rho_a+p_a)} k^2 \Psi^{(0)} d\eta \sim
\nonumber \\
&& c_1(k)\eta^{-2}\left[ \eta^{1-2q}- 1  \right].
\end{eqnarray}
Every integral gives a time-dependent contribution and a constant
one (taking also into account pre-bounce and Bounce contributions,
which are of the same order).

As a second step, we calculate the integrals using $\Psi^{(1)}$
and $\tilde{\zeta}^{(1)}$
\begin{eqnarray}
&\Psi^{(2)}=&\frac{\mathcal{H}}{a^2} \int
\frac{a^2}{\mathcal{H}^2} \tilde{\zeta}^{(1)} d\eta \sim\nonumber
\\
&&
c_1(k)\eta^{-1-2q} \left[ \eta^{2}- \eta^{1+2q}-1  \right] \\
&\tilde{\zeta}^{(2)}=&- a^2(\rho_a+p_a) \int
\frac{c_a^2\mathcal{H}}{a^2(\rho_a+p_a)} k^2 \Psi^{(1)} d\eta \sim
\nonumber \\
&& c_2(k)k^2\eta^{-2}\left[ \eta^2-\eta^{1-2q}- 1  \right].
\end{eqnarray}
In the same way we can build $\Psi^{(3)}$ and
$\tilde{\zeta}^{(3)}$, and so on. The complete solution is the sum
of all the partial contributions
\begin{eqnarray}
&& \Psi=\sum\limits_{i=0}^{\infty} \Psi^{(i)} \\
&& \tilde{\zeta}=\sum\limits_{i=0}^{\infty} \tilde{\zeta}^{(i)}.
\end{eqnarray}

However, since we are interested in the solution just before
horizon re-entry, we can drop all terms that are just $k^2\eta^2$
times some other term already present in the series. In fact,
these terms determine the first oscillation of the mode at horizon
re-entry. For example, the first term in the brackets of
$\Psi^{(2)}$ and $\tilde{\zeta}^{(2)}$ are just $k^2\eta^2$ times
$\Psi^{(0)}$ and $\tilde{\zeta}^{(0)}$ respectively. The last
terms in the same brackets are just $k^2$ times $\Psi^{(0)}$ and
$\tilde{\zeta}^{(0)}$ respectively, so that they are just $k^2$
corrections to the homogeneous modes (remember that we are
studying modes with $k \ll 1$). In this simplified example, all
$\Psi^{(i)}$ and $\tilde{\zeta}^{(i)}$ with $i \geq 3$ give
subleading terms according to this prescription. We are finally
left with the following post-bounce solution
\begin{eqnarray}
&\Psi \sim &k^{-\frac{3}{2}-q}\eta^{-1-2q} +k^{-\frac{1}{2}+q}+\nonumber \\
&&k^{-\frac{1}{2}+q}\eta^{-1-2q} + k^{\frac{1}{2}-q} \label{Psipostsimple}\\
&\tilde{\zeta} \sim & k^{-\frac{1}{2}+q}\eta^{-2} +
k^{\frac{1}{2}-q}\eta^{-1-2q}+
\nonumber \\
&&k^{\frac{1}{2}-q}\eta^{-2}+k^{\frac{3}{2}+q}\eta^{-1-2q},
\end{eqnarray}
where we have replaced $c_i(k)$ by their explicit spectral
dependence.

For each variable, the first term is the homogeneous mode, the
second and the third come from the integral of the homogeneous
mode of the other variable, the fourth term is the only new term
coming from the second recursion. Now we can discuss the evolution
of these four modes in the post-bounce. We can distinguish two
cases: fast contraction--expansion ($q>1/2$) and slow
contraction--expansion ($q<1/2$).

Let us start by the case $q>1/2$. Just after the bounce ($\eta
\sim 1$), the dominant mode is always the reddest one. We see that
$\Psi$ is by far dominated by its homogeneous mode, while
$\tilde{\zeta}$ is dominated by the integral of $\Psi$. However,
when we evaluate the modes at horizon re-entry ($\eta \sim 1/k$),
$\tilde{\zeta}$ is still dominated by the third mode, which comes
from the integral of the homogeneous mode of $\Psi$ and in
particular is the contribution coming from regions close to the
bounce. On the contrary, at $\eta \sim 1/k$ the homogeneous mode
of $\Psi$ has decayed below the fourth mode in Eq.
(\ref{Psipostsimple}). This constant mode comes from the integral
of the dominant mode of $\tilde{\zeta}$. This mode has the
original spectrum of the homogeneous mode of $\Psi$ multiplied by
$k^2$, which is typically blue.

In the case of slow evolution ($q<1/2$), both $\Psi$ and
$\tilde{\zeta}$ are dominated by their homogeneous modes at the
bounce ($\eta \sim 1$); $\tilde{\zeta}$ does not acquire a mode
dominating its homogeneous mode and, consequently, the fourth mode
of $\Psi$ remains subdominant till re-entry. In this case, the
dominant mode in $\Psi$ at re-entry is still the homogeneous
decaying mode, endowed with its red spectrum, which becomes
scale-invariant in the limit $q\rightarrow 0$. Of course, this
situation cannot be compatible with observations, since the
potentially scale-invariant spectrum is carried by a decaying mode
that becomes subdominant with respect to a constant mode just at
horizon re-entry. Moreover, as we shall see later, the interplay
with the secondary component (which we have neglected in this
simplified treatment) is essential in this regime, since it
provides those isocurvature modes that dominate at horizon
re-entry.

\begin{figure}
\resizebox{\hsize}{!}{\includegraphics{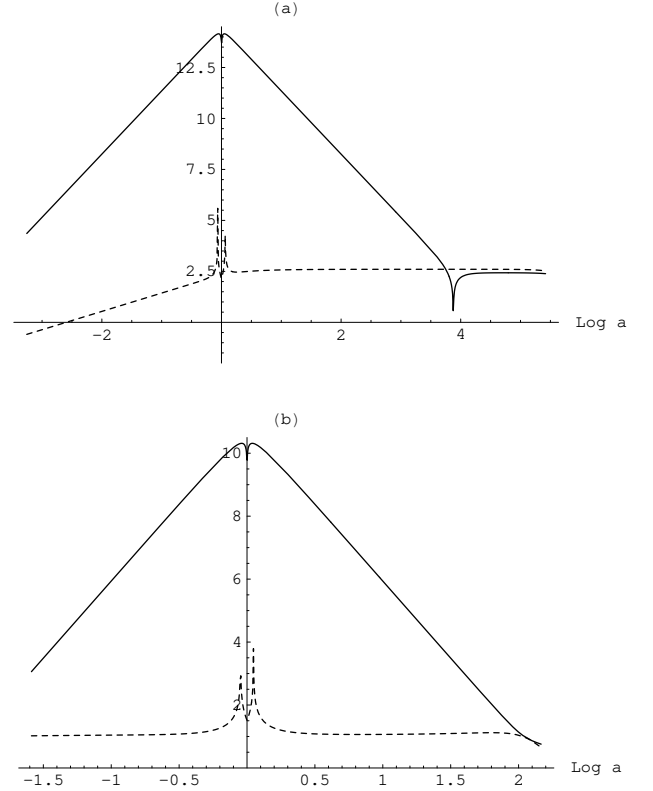}} \caption{(a)
Simplified perturbations in a bouncing cosmology with $\alpha=0$,
$c_a^2=0.4$, $c_b^2=2.5$. The solid line is $\mathrm{Log} |\Psi|$
and the dashed line is $\mathrm{Log} |\zeta|$. (b) The same as (a)
with $c_a^2=1.6$.
 }
 \label{Fig Simple}
\end{figure}

The two cases of this simplified example are synthesized in Fig.
\ref{Fig Simple}, where we have plotted $\Psi$ and $\zeta$ (which
is more familiar than $\tilde{\zeta}$ to people working in
cosmological perturbations). Following \cite{AllWan}, we have
chosen $\mathrm{Log}\; a(\eta)$ as time coordinate. We see that in
Fig. \ref{Fig Simple}a, where we have chosen a model with $q>1/2$,
$\zeta$ grows in the pre-bounce, dominated by the integral of
$\Psi$. Then it sets to a constant value in the post-bounce. The
Bardeen potential pre-bounce growing mode turns into a decaying
mode in the post-bounce, which becomes subdominant when $\Psi$
becomes of the order of $\zeta$. In Fig. \ref{Fig Simple}b, where
we have chosen a model with $q<1/2$, $\zeta \simeq \eta^2
\tilde{\zeta}$ is dominated by its homogeneous (constant) mode in
the pre-bounce and in the post-bounce. The decaying mode of $\Psi$
becomes of the same order as the constant mode just at horizon
re-entry. In this case, for very small $q$, we have an almost
scale-invariant spectrum for the decaying mode of $\Psi$, while
its constant mode has a blue spectrum.

Notice also the presence of two spikes in $\zeta$ before and after
the bounce. These are the mentioned divergences at the NEC
violation times, which make $\zeta$ unsuitable for the description
of the evolution of scalar perturbations close to the bounce.

\section{Perturbations at the onset of horizon re-entry}

In this section, we apply the method explained in the previous
section to the whole system of integral equations governing our
problem. We shall thus build the full solution of scalar
perturbations in the post-bounce, keeping all the relevant modes.
We shall then find out the dominant modes at the onset of horizon
re-entry for all gauge-invariant variables.

After four recursions in the integrations of Eqs.
(\ref{Psiint})--(\ref{zbint}), no new modes are added to the
post-bounce solution. Its final form reads

\begin{eqnarray}
&\Psi\sim & k^{\nu_a-2}\eta^{-1-2q} + k^{\nu_a}+ \nonumber\\
&&C_{a}k^{-\nu_a}\eta^{1-2q}+k^{-\nu_a}+\nonumber\\
&&C_{a}k^{-\nu_b}\eta^{1-2q}+X_{a} k^{-\nu_b}+\nonumber\\
&&C_{a}k^{\nu_b}\eta^{1-2q}+C_{a}k^{\nu_b}, \label{Psipost}
\end{eqnarray}
\begin{eqnarray}
&\tilde{\zeta}\sim &
k^{\nu_a}\eta^{-2}+k^{\nu_a}\eta^{-1-2q}+\nonumber\\
&&k^{-\nu_a}\eta^{-2}+C_{a}k^{-\nu_a}\eta^{-1-2q}+\nonumber\\
&&X_{a}k^{-\nu_b}\eta^{-2}+C_{a}k^{-\nu_b}\eta^{-1-2q}+\nonumber\\
&&C_{a}k^{\nu_b}\eta^{-2}+C_{a}k^{\nu_b}\eta^{-1-2q},
\label{zetapost}
\end{eqnarray}
\begin{eqnarray}
&\Psi_b \sim & k^{\nu_a}\eta^{-1-q(1+3c_b^2)}+X_{a}k^{\nu_a+2}\eta^{-2q}+\nonumber\\
&& k^{-\nu_a+2}\eta^{1-q(1+3c_b^2)}+ k^{-\nu_a+2}\eta^{-2q}+\nonumber\\
&& k^{-\nu_b}\eta^{-1-q(1+3c_b^2)}+ k^{-\nu_b+2}\eta^{-2q}+\nonumber\\
&& k^{\nu_b+2}\eta^{1-q(1+3c_b^2)}+
k^{\nu_b+2}\eta^{-2q},\label{Psibpost}
\end{eqnarray}
\begin{eqnarray}
& \tilde{\zeta}_b \sim &
X_{a}k^{\nu_a}\eta^{-1-2q}+k^{\nu_a}\eta^{-q(1+3c_b^2)}+\nonumber\\
&&k^{-\nu_a}\eta^{-1-2q}+k^{-\nu_a}\eta^{-q(1+3c_b^2)}+\nonumber\\
&&k^{-\nu_b}\eta^{-1-2q}+k^{-\nu_b}\eta^{-q(1+3c_b^2)}+\nonumber\\
&&k^{\nu_b}\eta^{-1-2q}+C_{a}
k^{\nu_b}\eta^{-q(1+3c_b^2)}.\label{zbpost}
\end{eqnarray}

For each mode, we have just written their $k$ dependence, given by
the constants $c_i(k)$, and their $\eta$ dependence, discarding
any numerical factor. As for the pre-bounce solution, we have used
the factors $C_a$ and $X_a$ to include the case $\alpha=0$. Let us
explain how these factors arise. The coefficients $C_{a}$ come
because of the integration of $\tilde{\zeta}_b$ in Eq.
(\ref{zetaint}). If $\alpha$ vanishes, this term is absent and the
same modes come only through integration of $\Psi_b$, which
contain, however, an extra $k^2\eta^2$ factor. Finally, the
$X_{a}$ coefficients come from subtle cancellations in the
solution for two perfect fluids $\alpha_a=0$, which require some
more discussion.

The first cancellation is in the integral of $\Psi_b$ in Eq.
(\ref{zetaint}). In fact, the integrand is an odd function of the
background. In this situation, the pre-bounce, the Bounce and the
post-bounce contributions cancel exactly, leaving no trace except
for a decaying mode in the post-bounce. Successive recursions do
not involve odd functions. The terms reappear multiplied by a
$k^2$, but with the same time dependence. This determines the
$X_{a}$ factors in Eqs. (\ref{Psipost}) and (\ref{zetapost}).

The second cancellation involves several integrals and can only be
demonstrated using the exact solution of the two-fluid background
(available in terms of hypergeometric functions). One has to pick
out the integrals of $\Psi$ in $\Psi_b$ and $\tilde{\zeta}$ (the
latter corrected by the contribution of $\Psi_b$ to
$\tilde{\zeta}$) and plug them into Eq. (\ref{zbint}), where the
cancellation occurs. This explains the $X_{a}$ in Eqs.
(\ref{Psibpost}) and (\ref{zbpost}).

Comparing Eqs. (\ref{Psipost})--(\ref{zbpost}) to Eqs.
(\ref{PsiPre})--(\ref{zbPre}), we see that each mode in the
pre-bounce gives rise to two modes in the post-bounce, one with
the time-dependence of the homogeneous mode and one with the time
dependence of the integral of the partner homogeneous mode. In
some cases, these modes may be multiplied by $k^2\eta^2$ or $k^2$
factors.

As in the simplified example of the previous section, also here
$\Psi$ has no constant mode with a possibly scale-invariant
spectrum. At most, if there are no other modes dominating, the
scale-invariant spectrum is carried by a decaying mode. This is
hard to reconcile with experimental data. Nevertheless, we shall
prove that there is always another dominant mode.

\subsection{Discussion of the case $\alpha\neq 0$}
\label{Sect. alphanonzero}

All modes essentially depend on the characteristics of the
background, which are encoded in the 3 parameters $c_a^2$,
$\Gamma_a$ (or equivalently $q=2/(1+3\Gamma_a)$),
$c_b^2=\Gamma_b$. However, $c_a^2$ comes always as a numerical
coefficient and never enters the exponents of $k$ and $\eta$.
Following our policy of discarding all numerical factors, we can
discuss our results in terms of the remaining two parameters.

Of course, not all values of $\Gamma_a$ and $c_b^2$ are relevant
to bouncing cosmologies. In fact, in order to have a bounce, the
ratio $\rho_b/\rho_a$ must grow during contraction. This requires
$c_b^2>\Gamma_a$. Moreover, $\Gamma_a$ should be greater than
$-1/3$, in order to keep $q>0$ and have an ordinary contraction.
Finally, $c_b^2$ must be greater than zero for a correct vacuum
normalization. The allowed region is shown in Fig. \ref{Fig
allow}.

\begin{figure}
\resizebox{\hsize}{!}{\includegraphics{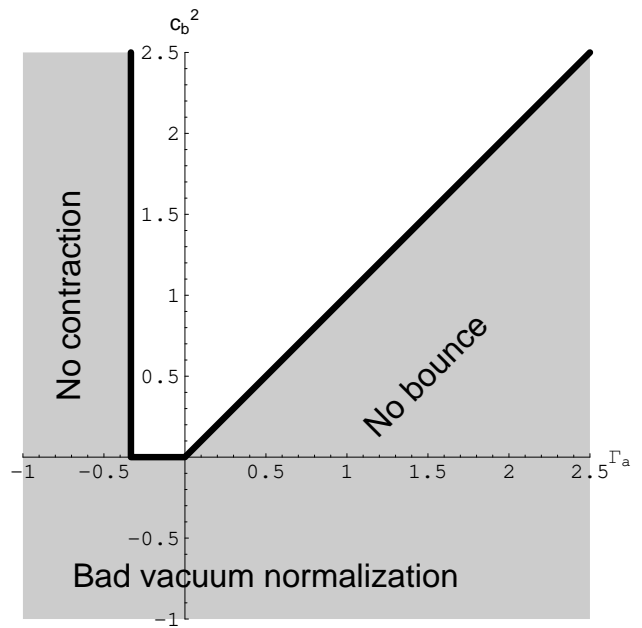}} \caption{In
the plane ($\Gamma_a$,$c_b^2$) we show the interesting region for
bouncing cosmologies in the case $\alpha \neq 0$. The shaded
regions are excluded as giving no initial contraction, no bounce,
or bad vacuum normalization of the secondary component.
 }
 \label{Fig allow}
\end{figure}

In principle, we should also exclude all values of $c_b^2>1$,
since they cause fluctuations in the secondary component to
propagate faster than light. However, since we are mainly
interested in the effects of the kinematics of the bounce on the
perturbations, in order to explore the slow evolution region
$q<1/2$, we shall also include the region $c_b^2>1$ in our
analysis.

\begin{figure}
\resizebox{\hsize}{!}{\includegraphics{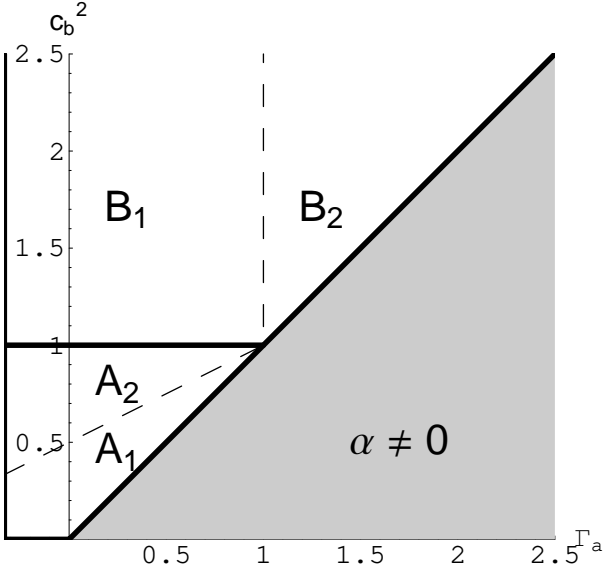}}
\caption{ In the case $\alpha\neq 0$, we have four regions. In
regions $A_1$, $A_2$ the spectrum of the dominant component
prevails in $\Psi$, while in regions $B_1$, $B_2$, the spectrum of
the secondary component prevails.
 }
 \label{Fig alphanonzero}
\end{figure}

Comparing all the modes in Eqs. (\ref{Psipost})--(\ref{zbpost}),
we can distinguish four regions depending on the mode dominating
each variable. These regions, shown in Fig. \ref{Fig
alphanonzero}, and their dominant modes are
\begin{eqnarray}
&A_1:& \Psi\sim k^{\nu_a}, \; \tilde{\zeta} \sim k^{\nu_a}\eta^{-2},\nonumber \\
&& \Psi_b \sim k^{\nu_a}\eta^{-1-q(1+3c_b^2)}, \; \tilde{\zeta}_b
\sim k^{\nu_a}\eta^{-q(1+3c_b^2)} \\
&A_2:& \Psi\sim k^{\nu_a}, \; \tilde{\zeta} \sim k^{\nu_a}\eta^{-2},\nonumber \\
&& \Psi_b \sim k^{\nu_a+2}\eta^{-2q}, \; \tilde{\zeta}_b
\sim k^{\nu_a}\eta^{-1-2q} \label{A2}\\
&B_1:& \Psi\sim k^{-\nu_b}, \; \tilde{\zeta} \sim k^{-\nu_b}\eta^{-2},\nonumber \\
&& \Psi_b\sim  k^{-\nu_b+2}\eta^{-2q}, \; \tilde{\zeta}_b
\sim k^{-\nu_b}\eta^{-1-2q} \label{B1}\\
&B_2:& \Psi\sim k^{-\nu_b}\eta^{1-2q}, \; \tilde{\zeta} \sim k^{-\nu_b}\eta^{-1-2q},\nonumber \\
&& \Psi_b \sim k^{-\nu_b+2}\eta^{-2q}, \; \tilde{\zeta}_b \sim
k^{-\nu_b}\eta^{-1-2q}.
\end{eqnarray}
The borders between the different regions are
\begin{eqnarray}
&A_1-A_2: & c_b^2=\frac{1+\Gamma_a}{2}\\
&A_2-B_1: & c_b^2=1\\
&B_1-B_2: & \Gamma_a=1.
\end{eqnarray}

We see that $\Psi$ is dominated by the spectrum of the primary
component for all bouncing cosmologies where the secondary
component has $c_b^2<1$ (regions $A_1$ and $A_2$). Perturbations
are also adiabatic, since $\zeta \sim \eta^2 \tilde{\zeta}$ is
constant outside the horizon. On the contrary, for $c_b^2>1$,
including also slowly evolving cosmologies ($\Gamma_a>1$), $\Psi$
is dominated by the spectrum of the secondary component. For
$\Gamma_a>1$ (region $B_2$), this mode is isocurvature, while for
$\Gamma_a<1$ (region $B_1$), this mode is adiabatic. The
distinction between regions $A_1$ and $A_2$ comes from the
different time dependence in $\Psi_b$ and $\tilde{\zeta}_b$.

Allen \& Wands studied a bounce where the principal component is a
scalar field with a tracking potential, so that $c_a^2=1$, but
$\alpha_a \neq 0$ and $\Gamma_a$ can assume any value between
$-1/3$ and 0. Their secondary component is a free ghost scalar
field ($c_b^2=1$, $\alpha_b=0$). Their case thus lies just on the
border line between region $A_2$ and $B_1$. From Eqs. (\ref{nua})
and (\ref{nub}), $\nu_a=\nu_b$ and the spectra of the two
components coincide. They find that the final form of $\Psi$,
after the homogeneous mode has decayed, is the one given in Eqs.
(\ref{A2}) or (\ref{B1}).

\subsection{Discussion of the case $\alpha=0$} \label{Sect. alphazero}

The case $\Gamma_a=c_a^2$ and $\Gamma_b=c_b^2$ contains all
bounces developed by two perfect fluids or free scalar fields. In
this case it is possible to write analytic solutions of the
background in terms of hypergeometric functions, which are very
useful to check for particular cancellations of modes, as
explained before. In fact, all bounces described in this class
enjoy time-reversal symmetry and thus the integrals of odd
functions vanish. Therefore, the classification presented in the
previous subsection has to be revised in order to take into
account these cancellations. Furthermore, we cannot accept values
$\Gamma_a=c_a^2<0$, for vacuum normalization reasons, and we are
restricted to the upper half of the first quadrant.

\begin{figure}
\resizebox{\hsize}{!}{\includegraphics{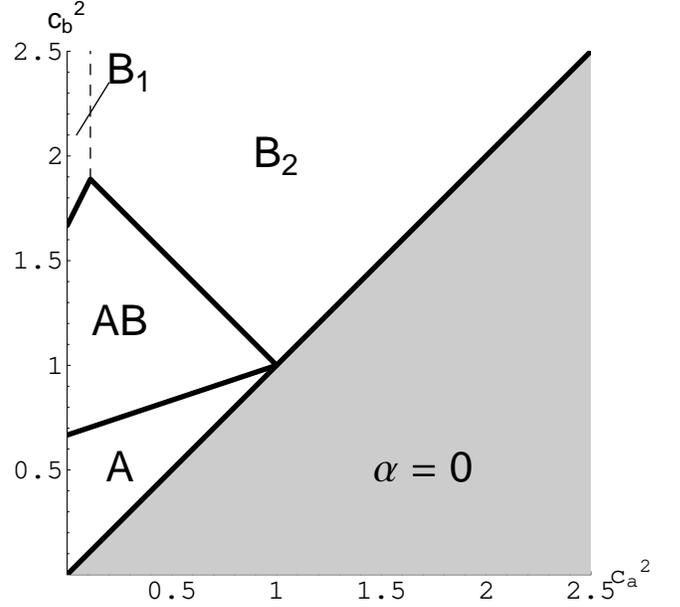}} \caption{
In the case $\alpha=0$, we have four regions. In regions $A$ and
$AB$ the spectrum of the dominant component prevails in $\Psi$,
while in regions $B_1$, $B_2$, the spectrum of the secondary
component prevails.
 }
 \label{Fig alphazero}
\end{figure}

Four different regions show up. They are illustrated in Fig.
\ref{Fig alphazero} and are characterized by the following modes
\begin{eqnarray}
&A:& \Psi\sim k^{\nu_a}, \; \tilde{\zeta} \sim k^{\nu_a}\eta^{-2},\nonumber \\
&& \Psi_b \sim k^{\nu_a}\eta^{-1-q(1+3c_b^2)}, \; \tilde{\zeta}_b
\sim k^{\nu_a}\eta^{-q(1+3c_b^2)} \\
&AB:& \Psi\sim k^{\nu_a}, \; \tilde{\zeta} \sim k^{\nu_a}\eta^{-2},\nonumber \\
&& \Psi_b \sim k^{-\nu_b+2}\eta^{-2q}, \; \tilde{\zeta}_b
\sim k^{-\nu_b}\eta^{-1-2q} \\
&B_1:& \Psi\sim k^{-\nu_b+2}, \; \tilde{\zeta} \sim k^{-\nu_b+2}\eta^{-2},\nonumber \\
&& \Psi_b \sim k^{-\nu_b+2}\eta^{-2q}, \; \tilde{\zeta}_b
\sim k^{-\nu_b}\eta^{-1-2q}\\
&B_2:& \Psi\sim k^{-\nu_b+2}\eta^{3-2q}, \; \tilde{\zeta} \sim k^{-\nu_b+2}\eta^{1-2q},\nonumber \\
&& \Psi_b \sim k^{-\nu_b+2}\eta^{-2q}, \; \tilde{\zeta}_b \sim
k^{-\nu_b}\eta^{-1-2q}.
\end{eqnarray}
The borders between the different regions are
\begin{eqnarray}
&A-AB: & c_b^2=\frac{2+c_a^2}{3}\\
&AB-B_1: & c_b^2=\frac{5}{3}+2c_a^2\\
&AB-B_2: & c_b^2=2-c_a^2\\
&B_1-B_2: & c_a^2=\frac{1}{9}.
\end{eqnarray}

The region where $\Psi$ is dominated by the spectrum of the first
component is enlarged towards $c_b^2>1$. This include regions $A$
and $AB$. In regions $B_1$ and $B_2$, $\Psi$ gets the spectrum of
the secondary component multiplied by $k^2$. In $B_1$,
perturbations are adiabatic and in $B_2$ they are isocurvature. It
is interesting to note how region $B_1$ has been reduced by the
cancellation discussed above. Only in region $A$ do
$\tilde{\zeta}_b$ and $\Psi_b$ take on the spectrum of the
dominant component, while in all other regions they keep the
original spectrum multiplied by $k^2$. In region $AB$, therefore,
we have coexistence of the two spectra in $\tilde{\zeta}_a$ and
$\tilde{\zeta}_b$, though $\tilde{\zeta}$ is dominated by
$\tilde{\zeta}_a$. This is an effect of the other cancellation
discussed before.

\section{Numerical solutions}

A numerical test of the above predictions has been performed by
integrating the set of differential equations
(\ref{EqPsi})--(\ref{Eqzb}) using Mathematica. Initial conditions
have been posed far from the bounce using the vacuum normalization
prescription described in Sect. \ref{Sect. vacuum}. The default
Mathematica method for resolution of ordinary differential
equations (which is a combination of Gear and Adams methods)
cannot go through the bounce, because the step size falls to
zero asymptotically as $\eta \rightarrow 0$. This is a consequence
of the presence of $\mathcal{H}^{-1}$ factors in the differential
equations, which requires a high precision cancellation in the
numerators. However, all variables and their derivatives have
regular limits at the bounce, as can be explicitly checked on the
so-obtained pre-bounce solutions. Thus a fixed step size method
would overcome this problem. In fact, a standard Runge-Kutta
method nicely follows the evolution of perturbations in any phases
and in particular across the bounce. However, this method is too slow to be
employed for an extensive investigation of the parameter space.
As a good compromise, we use the default Mathematica method
far from the bounce, where it is faster than Runge-Kutta and as
reliable. Only in the interval  $|\eta|<10^{-5}$
we switch to a Runge-Kutta method which allows us to cross
the bounce. After the bounce we return to the default Mathematica
method until horizon re-entry. The combination of the two methods
then gives a very efficient and reliable numerical integrator for
perturbations in bouncing cosmologies.

Since we are mainly interested in extracting the dominant mode at
horizon re-entry, we have followed the evolution of two solutions
with different $k$ and we have evaluated the spectrum of the
generic variable $f$ in the set
$\{\Psi,\tilde{\zeta},\Psi_b,\tilde{\zeta}_b\}$ as
\begin{equation}
\frac{\log f_{k_1}(\eta_{re})-\log f_{k_2} (\eta_{re})}{\log k_1 -
\log k_2},
\end{equation}
where the subscripts 1 and 2 refer to the two solutions with
different values of $k$, and $\eta_{re}$ has been chosen of the
order of $1/k$ but small enough to avoid contamination from the
first oscillation. Of course, using more solutions would have
given a more accurate and reliable fit of the spectrum, but it
turns out that the accuracy of the simulation allows this simple
and fast choice.

The other important prediction is on the time-dependence of the
dominant mode, i.e. the exponent of $\eta$ (hereafter we shall
call it time index). This can be easily extracted as
\begin{equation}
\left. \frac{d \log f}{d\log \eta}\right|_{\eta=\eta_{re}}.
\end{equation}

We have tested two cosmological models: a bounce with two perfect
fluids, and a bounce with a scalar field endowed with a tracking
potential plus a perfect fluid. In both cases the secondary fluid
has negative energy density in order to trigger the bounce. We
shall present the results separately.

\subsection{Numerical testing of the two-perfect-fluid model}

With this class of numerical simulations we can test the
predictions of Sect. \ref{Sect. alphazero}. We have spanned the
plane $(c_a^2,c_b^2)$ starting from $c_a^2=0.1$ up to $c_a^2=2$
with $c_b^2$ running from $c_a^2+0.1$ up to $c_b^2=2.5$, with
steps of 0.1 in both parameters.

\begin{figure*}
\resizebox{11cm}{!}{\includegraphics{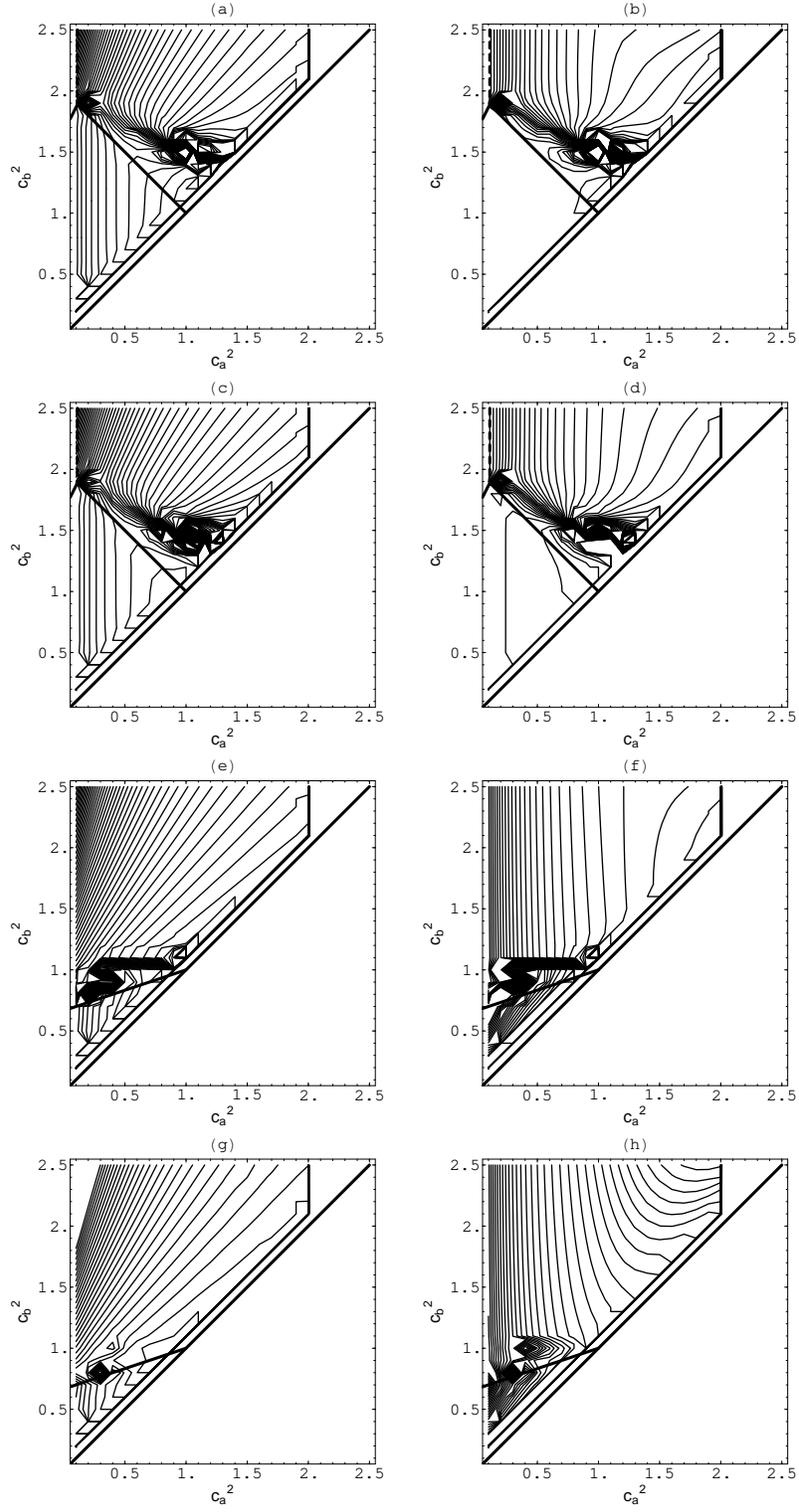}} \caption{
On the left panels we have the contour plots of the spectral
indices and on the right panels the time indices in the case
$\alpha_a=\alpha_b=0$. (a) and (b) refer to $\Psi$, (c) and (d) to
$\tilde{\zeta}$, (e) and (f) to $\Psi_b$, (g) and (h) to
$\tilde{\zeta}_b$.
 }
 \label{Fig 2Dalphazero}
\end{figure*}

After the evaluation of the spectral and time index for all
variables, just to have a global view of the parameter plane, we
can draw contour plots for each index and compare these with the
diagram of Fig. \ref{Fig alphazero}. These are shown in Fig.
\ref{Fig 2Dalphazero}.

The orientation of the contours in the plots of Fig. \ref{Fig
2Dalphazero}a-d evidently shows the existence of two main regions
of the parameter plane where the spectrum and the time dependence
of the dominant mode in $\Psi$ and $\tilde{\zeta}$ at re-entry are
different. The calculated border line between region $AB$ and
region $B_2$ is superimposed in these figures. Actually, the
numerical border line seems to be slightly higher towards larger
values of $c_b^2$. However, this border line is very sensitive to
the choice of the evaluation time $\eta_{re}$. Taking different
times we can see that the numerical border line tends to the
theoretical one in the limit $\eta_{re}\sim 1/k$. In practice, one
should keep in mind that close to the border line between regions
$AB$ and $B_2$ the two competing modes are comparable, giving rise
to mixed initial conditions for the evolution inside the horizon.
This also provokes glitches in the evaluation of the indices in
these intermediate cases, which show up in our plots with a large
density of contours on the upper side of the border line.

In Fig. \ref{Fig 2Dalphazero}e-h, the contours of spectral and
time indices of $\Psi_b$ and $\tilde{\zeta}_b$ are represented
together with the border line of regions $A$ and $AB$. With the
same {\it caveat } as before, also this theoretical border line is
respected.

\begin{figure*}
\resizebox{11cm}{!}{\includegraphics{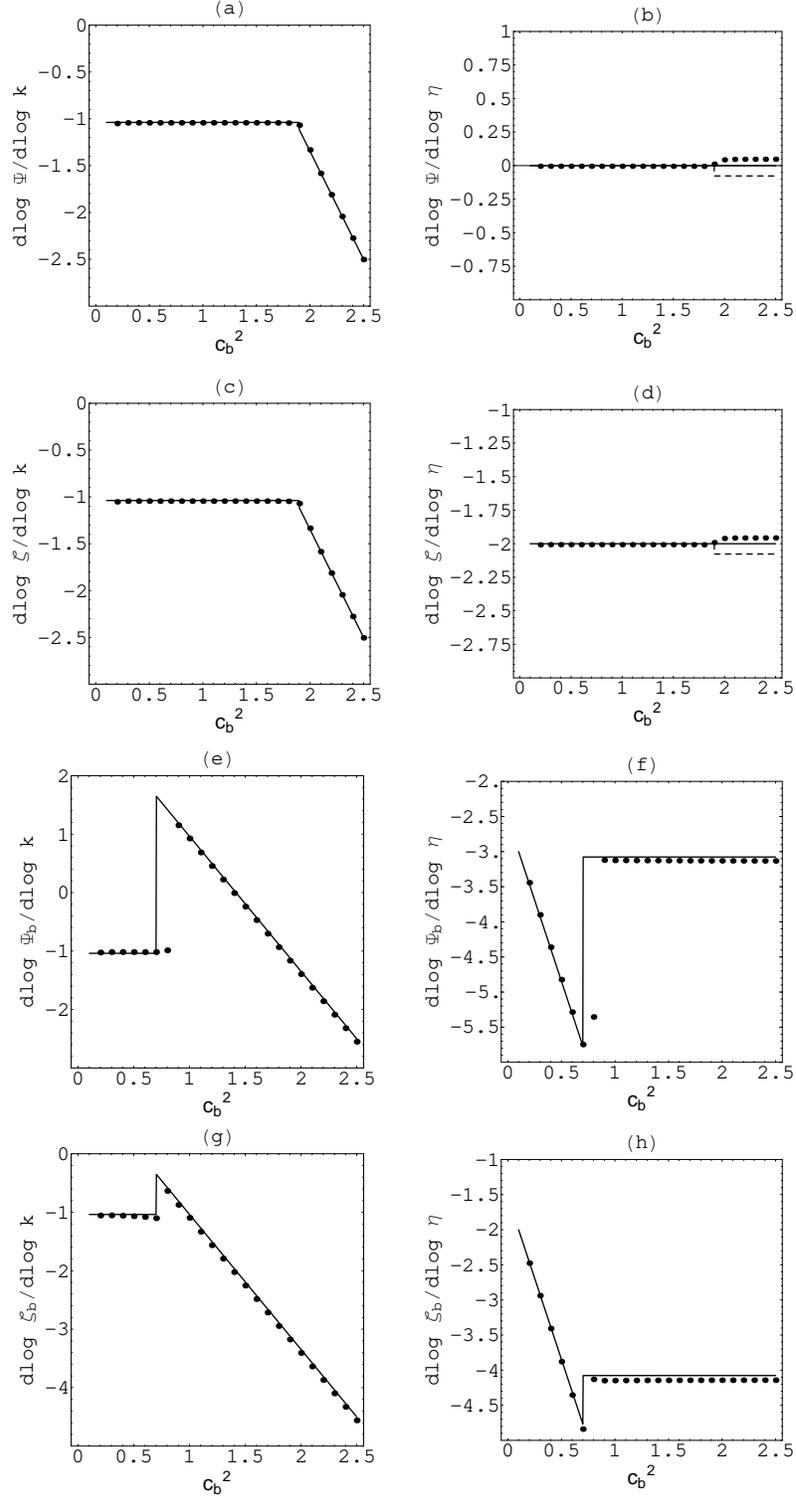}} \caption{
On the left panels we have the spectral indices and on the right
the time indices for our variables in the case $\alpha=0$, keeping
$c_a^2=0.1$ and varying $c_b^2$. (a) and (b) refer to $\Psi$, (c)
and (d) to $\tilde{\zeta}$, (e) and (f) to $\Psi_b$, (g) and (h)
to $\tilde{\zeta}_b$.
 }
 \label{Fig Spettrizero}
\end{figure*}

Now, let us make a closer comparison between the theoretical
values of the indices and those found numerically. In Fig.
\ref{Fig Spettrizero} we represent the results for a section of
the plane $(c_a^2,c_b^2)$ at fixed $c_a^2=0.1$ and varying
$c_b^2$. This line crosses three of the four theoretical regions,
namely regions $A$, $AB$ and $B_1$. The numerical values are
plotted along with the lines representing the theoretical
behaviour of the indices. We can see that the agreement is very
satisfactory for all gauge-invariant variables.

The spectra of $\Psi$ and $\tilde{\zeta}$ are perfectly reproduced
both in regions $A$, $AB$ and in region $B_1$, where they are
dominated by the spectrum of the secondary fluid. The distinction
between regions $B_1$ and $B_2$ is only in the time indices of
$\Psi$ and $\tilde{\zeta}$. In Figs. \ref{Fig Spettrizero}b,d we
have also plotted as a dashed line the theoretical behaviour that
these variables would have if region $B_1$ did not exist and there
were only region $B_2$. We see that while the numerical values
change but are still compatible with the expectation of region
$B_1$, they cannot be compatible with the behaviour predicted by
region $B_2$.

The secondary fluid variables clearly show the transition between
regions $A$ and $AB$. Only one point in $\Psi_c$ still seems to
follow the dominant mode of region $A$ rather than that of region
$AB$ for $c_b^2=0.7$. As already stated, this depends on a
conservative choice of $\eta_{re}$, which eliminates
contaminations from the oscillations but can sometimes select a
not-yet-decayed mode in almost marginal cases like this.

\begin{figure}
\resizebox{\hsize}{!}{\includegraphics{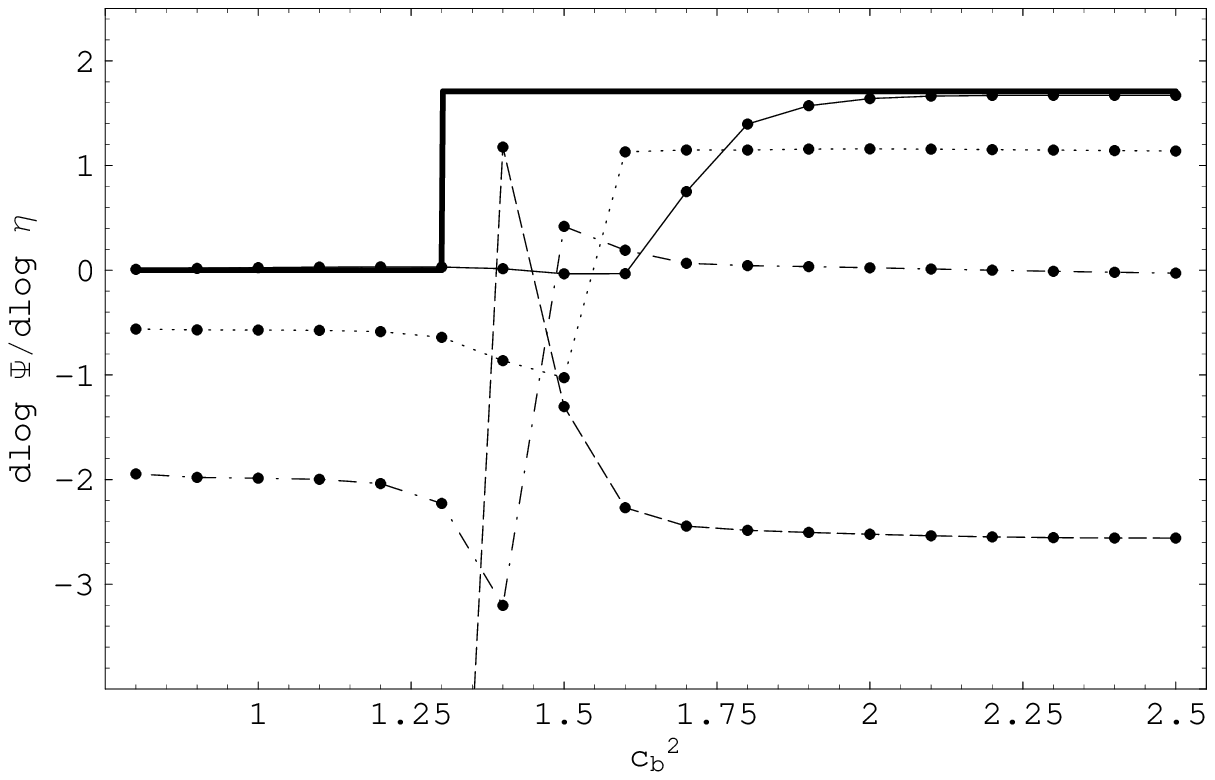}} \caption{ Time
index for $\Psi$ for $c_a^2=0.7$ and varying $c_b^2$. The
evaluation has been done at different $\eta_{re}$. Points along
the solid line are taken at $\eta_{re}=0.4/(c_a k)$. The dotted
line is obtained with $\eta_{re}=1.5/(c_a k)$, the dot-dashed line
with $\eta_{re}=2.5/(c_a k)$, and the dashed line with
$\eta_{re}=3.5/(c_a k)$.
 }
 \label{Fig Etare}
\end{figure}

The existence of several modes that are of the same order at
$\eta_{re}$ is particularly evident for $c_a^2 \sim 1$ and $c_b^2
\gtrsim 1$. In Fig. \ref{Fig Etare} we have plotted the numerical
values for the time index of $\Psi$ obtained for different choices
of $\eta_{re}$, along the line in the parameter plane with
$c_a^2=0.7$. A conservative choice such as $\eta_{re}=0.4/(c_a k)$
leads to very good agreement with the theoretical values far from
the border line, but the points with $c_b^2$ close to the border
line between regions $AB$ and $B_2$ tend to follow the behaviour
of region $AB$. However, pushing $\eta_{re}$ towards the first
oscillation gives the necessary time to these points to feel the
mode of region $B_2$ and align to the others. However, the first
oscillation globally alters all evaluations, which are displaced
more and more from the theoretical super-horizon values. This
discussion evidently shows that close to the border line the two
competing modes contribute to determine the initial oscillations
after horizon re-entry.

\subsection{Numerical testing of the scalar-field model}

In the second model of the bounce we have tested, the dominant
fluid is replaced by a scalar field with an exponential potential.
As shown in Ref. \cite{AllWan}, the exponent of the potential
determines $\Gamma_a$ in the range $[0;1]$. At the same time,
$c_a^2$ is fixed to 1. With this model, we can thus test the
theoretical predictions for the case $\alpha \neq 0$. However, the
$\Gamma_a>1$ part of the plane, which corresponds to region $B_2$
of Fig. \ref{Fig alphanonzero}, remains unexplored numerically.
One more limitation arises at small $c_b^2$, where viable bouncing
backgrounds cannot be obtained. A complete study of this class of
backgrounds has been done only in the case $c_b^2=1$
\cite{AllWan}, so that the reasons why the dynamics does not flow
in the same way for small $c_b^2$ remains unknown to us. However,
since we are mainly interested in perturbations, we content
ourselves with higher values of $c_b^2$. This will be sufficient
to distinguish regions $A_1$, $A_2$ and $B_1$, as we shall see.

It must also be said that in this case we have found it convenient
to switch on a different set of differential equations close to
the bounce. This set is the same as Eqs.
(\ref{EqPsi})--(\ref{Eqzb}) but with $a \leftrightarrow b$. In
practice, to reduce numerical errors close to the bounce it is
better to follow $\Psi$, $\tilde{\zeta}$, $\Psi_a$ and
$\tilde{\zeta}_a$. In this way, the whole numerical simulation is
split in four phases: far pre-bounce, near pre-bounce, near
post-bounce and far post-bounce.

\begin{figure*}
\resizebox{11cm}{!}{\includegraphics{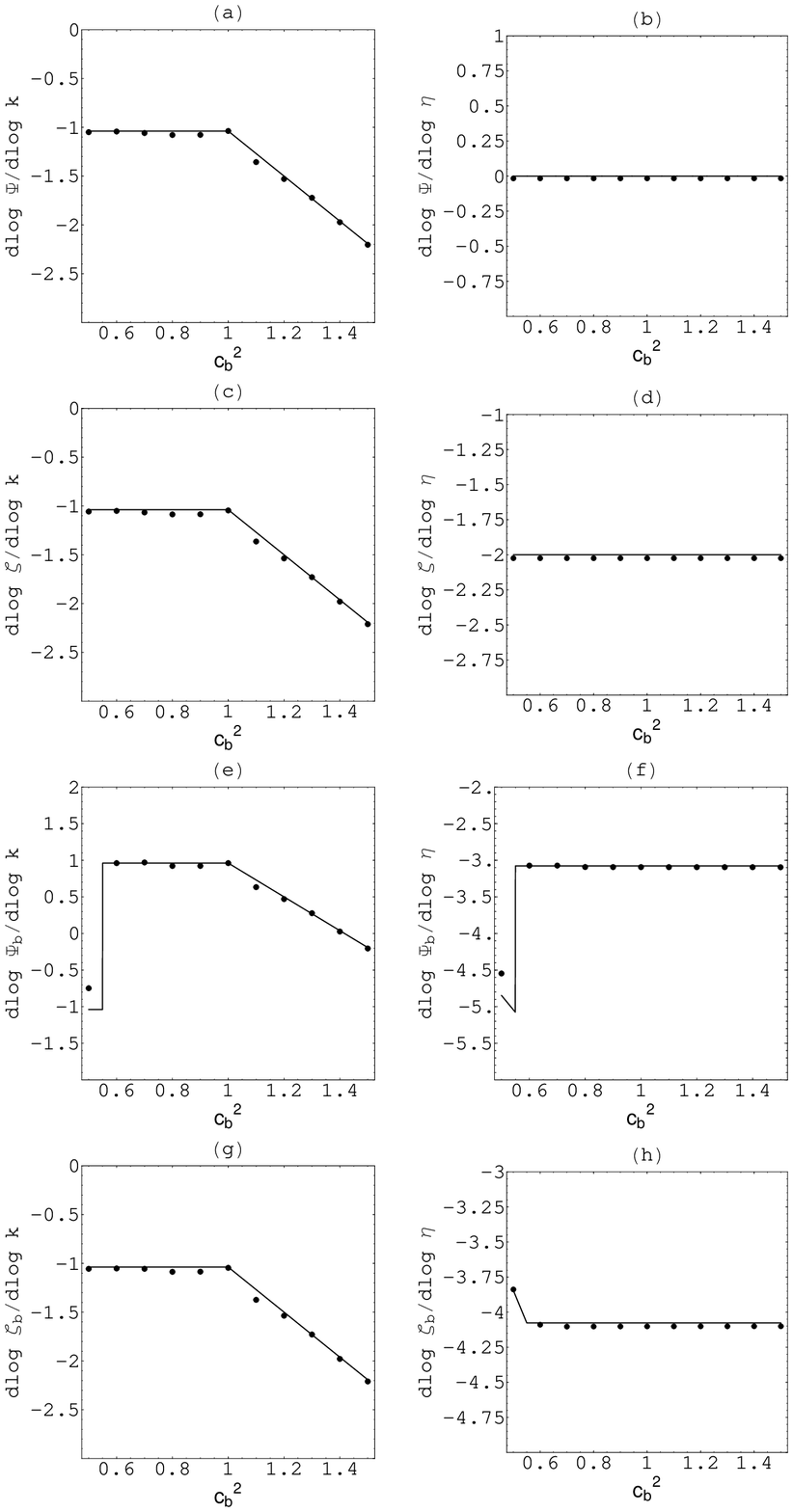}} \caption{
On the left panels we have the spectral indices and on the right
the time indices for our variables in the case $\alpha\neq 0$,
keeping $c_a^2=0.1$ and varying $c_b^2$. (a) and (b) refer to
$\Psi$, (c) and (d) to $\tilde{\zeta}$, (e) and (f) to $\Psi_b$,
(g) and (h) to $\tilde{\zeta}_b$.
 }
 \label{Fig Spettrialpha}
\end{figure*}

Figure \ref{Fig Spettrialpha} is the analogue of Fig. \ref{Fig
Spettrizero} for the bouncing cosmologies where the dominant
component is a scalar field. However, here, we have managed to
find good backgrounds only starting from $c_b^2=0.5$. In all the
spectra (Figs. \ref{Fig Spettrizero}a,c,e,g) the transition from
region $A_2$ to region $B_1$ at $c_b^2=1$, as predicted in Sect.
\ref{Sect. alphanonzero}, is evident. Only one point testifies the
existence of region $A_1$, where we see that spectrum and time
dependence of $\Psi_b$ and $\tilde{\zeta}_b$ change.

\section{Conclusions}

In this paper, we have addressed the problem of the evolution of
perturbations in bouncing cosmologies by testing explicit regular
models through the use of Einstein equations. At the moment, this
is the only effective and conclusive way to carry out a
theoretical investigation of the problem. In fact, in the absence
of clean model-independent general arguments supporting any claim,
the only answer can come from the study of regular bouncing models
within the frame of General Relativity, where the evolution of
perturbations can be followed explicitly from the initial state to
the final horizon re-entry in the current expansion era.

Within classical general relativity, a bounce from contraction to
expansion in spatially flat cosmologies can be realized only by
violating the NEC. All high-energy corrections to General
Relativity can thus be represented by a negative energy component
in the Friedman equations, which is important only in the bounce
phase. With this philosophy, we have completed a full study of a
very wide class of bouncing cosmologies, with two non-interacting
components, allowing the primary to have an intrinsic isocurvature
mode, in order to describe scalar fields and perfect fluids at the
same time. We have shown that the possibly scale-invariant
spectrum of $\Psi$ is carried by a homogeneous solution of the
differential equations, which decays in the post-bounce phase. No
other mode receives this spectrum, which is doomed to be only a
transient characteristic of the Bardeen potential. If we use
vacuum-normalized initial conditions for both components, we have
explicitly shown that this decaying mode becomes subdominant
w.r.t. some other mode before horizon re-entry. This general
result appears to contradict some claims made in the literature,
in particular those of Refs. \cite{PetPin} and \cite{Fin}.

As already explained in our previous paper \cite{BozVen}, in Ref.
\cite{PetPin}, Eq. (26) provides an analytical solution $f(\eta)=
\eta/(1+\eta^2)^2$ for perturbations outside the horizon, which is
valid both far from and through the bounce. However, the authors
use an approximate form ($f_{app}(\eta)=1/\eta^3$) of this
solution far from the bounce and then match $f_{app}$ to $f$ just
before the bounce. Had they used $f$ from the beginning throughout
the bounce, they would have found no mixing. No surprise,
therefore, that the mixing they find is at order $\eta^{-5}$,
which is just the next to leading order term in $f(\eta)$, which
is missing in $f_{app}(\eta)$. Had they kept the $\eta^{-5}$ term
in $f_{app}(\eta)$ they would have found mixing at the order
$\eta^{-7}$. We thus conclude that the mixing they find is just an
artifact of their matching procedure. We also note that a recent
paper \cite{GesBat} has criticized the splitting of $\Psi$
proposed by Ref. \cite{PetPin}, into two separate variables
$\Psi_+$ and $\Psi_-$ which satisfy the same equation of motion as
$\Psi$.

In the case of Ref. \cite{Fin}, the limit $\gamma \rightarrow 0$
($k \rightarrow 0$ in our notation), used to evaluate the final
spectrum, just selects the infrared modes at the bounce. However,
the final spectrum should be evaluated at the horizon re-entry,
after the decaying mode has gone away; it would correspond to
$\gamma \rightarrow 0$ while keeping $\gamma z$ (our $k\eta$) of
order 1.

The most important lesson from this class of models is probably
contained in the simplified example of Sect. \ref{Sect. simple}.
In fact, one may hypothesize any type of initial conditions for
the perturbations of the secondary component, but one should then
accept the basic facts, which are already included in the
decoupled evolution of the perturbations of the dominant
component. It should be concluded that in bouncing cosmologies
with fast evolution ($q>1/2$) the decaying mode of the Bardeen
potential is always subdominant with respect to its constant mode,
carrying a blue spectrum. In bouncing cosmologies with slow
evolution ($q<1/2$) the two modes are of the same order at horizon
re-entry. But then, even in the case when no other isocurvature
mode coming from coupling to other sources dominates, the
late-time phenomenology would be different from the observed one
\cite{AmeFin}.

There are at least three interesting directions in which  this
research may be extended.

The first is the study of more complicated models for
perturbations, going beyond the relations
(\ref{EOS})--(\ref{EOSb}). High-energy corrections may include
higher derivatives and anisotropic stress, which are definitely
outside the relations at the basis of our investigation. However,
a first negative result already comes from Ref. \cite{Car}, whose
conclusions are similar to ours.

The second direction is in allowing an interaction or a
correlation between the two components. A more realistic model,
also closer to the original ekpyrotic proposal, should include the
decay of the dominant component to radiation at the bounce. The
implementation of such models with a coherent treatment of
classical perturbation does not seem trivial.

Finally, the bounce integral may depend on an additional scale of
the theory, which is independent of the curvature scale of the
bounce. It is difficult to imagine a physical component with these
characteristics. However, the possibility remains open.

In conclusion, all these possible extensions must invoke some
special physics at the bounce in order to upset the basic result
that the scale-invariant spectrum of the Bardeen potential is
carried only by a decaying mode in the post-bounce. Up to now, no
explicit construction of such a possibility has been presented in
the literature and thus the claim that a scale-invariant spectrum
can be carried by a constant Bardeen potential after a bounce
remains an unproved conjecture.


\begin{thebibliography}{}

\bibitem{PBB} G. Veneziano, Phys. Lett. B {\bf 265}, 287 (1991);
 M. Gasperini and G. Veneziano, Astropart. Phys. {\bf 1}, 317 (1993);
R. Brustein, M. Gasperini, M. Giovannini, V. Mukhanov, and G.
Veneziano, Phys. Rev. D {\bf 51}, 6744 (1995);  M. Gasperini and
G. Veneziano, Phys. Rep. {\bf 373}, 1 (2003).

\bibitem{Ekp} J. Khoury, B.A. Ovrut, P.J. Steinhardt, and N. Turok, Phys.
Rev. D {\bf 64}, 123522 (2001) and Phys. Rev. D {\bf 66}, 046005
(2002); S. Gratton, J. Khoury, P. J. Steinhardt, and N. Turok,
Phys. Rev. D {\bf 69}, 103505 (2004); A. J. Tolley, N. Turok, and
P. J. Steinhardt, Phys. Rev. D {\bf 69}, 106005 (2004).

\bibitem{HorWit} P. Horava and E. Witten, Nucl. Phys. B {\bf 460}, 506 (1996) and B {\bf 475}, 94 (1996).

\bibitem{MFB} V.F. Mukhanov, H.A. Feldman, and R.H. Brandenberger,
Phys. Rep. {\bf 215}, 203 (1992).

\bibitem{Curv} K. Enqvist and M. S. Sloth, Nucl. Phys. B {\bf 626}, 395
(2002); D. H. Lyth and D. Wands, Phys. Lett. B {\bf 524}, 5
(2002).

\bibitem{Axion} V. Bozza, M. Gasperini, M. Giovannini, and G. Veneziano, Phys. Lett. B {\bf 543}, 14
(2002); Phys. Rev. D {\bf 67}, 063514 (2003).

\bibitem{Others}  D. Lyth, Phys. Lett. B {\bf 524}, 1 (2002);
R. Brandenberger and F. Finelli, JHEP {\bf 0111}, 056 (2001); D.
Lyth, Phys. Lett. B {\bf 526}, 173 (2002); J. Hwang, Phys. Rev. D
{\bf 65}, 063514 (2002); S. Tsujikawa, Phys. Lett. B {\bf 526},
179 (2002);  J. Martin, P. Peter, N. Pinto-Neto, and D.J. Schwarz,
Phys. Rev. D {\bf 65}, 123513 (2002); J. Hwang and H. Noh, Phys.
Lett. B {\bf 545}, 207 (2002); J. Martin, P. Peter, N. Pinto-Neto,
and D. J. Schwarz, Phys. Rev. D {\bf 67}, 028301 (2003); S.
Tsujikawa, R. Brandenberger, and F. Finelli, Phys. Rev. D {\bf
66}, 083513 (2002); C. Cartier, R. Durrer, and E. Copeland, Phys.
Rev. D {\bf 67}, 103517 (2003); P. Peter, N. Pinto-Neto, and D. A.
Gonzalez, JCAP {\bf 0312}, 003 (2003); P. Creminelli, A. Nicolis,
and M. Zaldarriaga, Phys. Rev. D {\bf 71}, 063505 (2005).

\bibitem{DerMuk} N. Deruelle and V.F. Mukhanov, Phys. Rev. D {\bf 52},
5549 (1995).

\bibitem{DurVer} R. Durrer and F. Vernizzi, Phys. Rev. D {\bf 66}, 083503
(2002).

\bibitem{K=1} J. Hwang and H. Noh, Phys. Rev. D {\bf 65}, 124010
(2002); C. Gordon and N. Turok, Phys. Rev. D {\bf 67}, 123508
(2003); J. Martin and P. Peter, Phys. Rev. D {\bf 68}, 103517
(2003); J. Martin and P. Peter, Phys. Rev. Lett. {\bf 92}, 061301
(2004);  N. Deruelle and A. Streich, Phys. Rev. D {\bf 70}, 103504
(2004); J. Martin and P. Peter, gr-qc/0406062.

\bibitem{PetPin} P. Peter and N. Pinto-Neto,
Phys. Rev. D {\bf 66}, 063509 (2002).

\bibitem{Fin} F. Finelli,
JCAP {\bf 0310}, 011 (2003).

\bibitem{Car} C. Cartier,
hep-th/0401036.

\bibitem{GGV} M. Gasperini, M. Giovannini and G. Veneziano,
Phys. Lett. B {\bf 569}, 113 (2003); Nucl. Phys. B {\bf 694}, 206
(2004).

\bibitem{AllWan} L. Allen and D. Wands,
Phys. Rev. D {\bf 70}, 063515 (2004).

\bibitem{BozVen} V. Bozza and G. Veneziano,  hep-th/0502047.

\bibitem{GesBat} G. Geshnizjani and T. J. Battefeld,
hep-th/0506139.

\bibitem{AmeFin} L. Amendola and F. Finelli, astro-ph/0411273.
\end{thebibliography}
\end{document}